\documentclass[preprint2]{exoplanet}
\usepackage{times}
\newcommand{\refs}{\par\noindent\hangindent=1pc\hangafter=1}

\newcommand{\AU}{\mbox{\textrm{AU}}}     
\newcommand{\yi}{\mbox{$\mathrm{yr}^{-1}$}}
\newcommand{\y}{\mbox{$\mathrm{yr}$}}

\newcommand{\si}{\mbox{$\mathrm{s^{-1}}$}}
\newcommand{\sdunits}{\mbox{$\mathrm{g}\,\mathrm{cm}^{-2}$}}
\newcommand{\kms}{\mbox{$\mathrm{km \, s^{-1}}$}}
\newcommand{\rp}{a} 
\begin{document}

\title{\textbf{\LARGE Planet Migration }\\
{\it To Appear in "Exoplanets," ed. S. Seager, Univ. Arizona Press.}}

\author {\textbf{\large Stephen H. Lubow}}
\affil{\small\em Space Telescope Science Institute}

\author {\textbf{\large Shigeru Ida}}
\affil{\small\em Tokyo Institute of Technology}

\begin{abstract}
\begin{list}{ } {\rightmargin 1in}
\baselineskip = 11pt
\parindent=1pc
{\small Planet migration is the process by which a planet's
orbital radius changes in time. 
The main agent for causing gas giant
planet migration is the gravitational interaction of the young
planet with the gaseous disk from which it forms. We describe
the migration rates resulting from these interactions based on
a simple model for disk properties.
These migration rates are higher than is reasonable for planet
survival. We discuss some proposed models for which the migration
rates are lower. There are major uncertainties in migration rates due to a lack
of knowledge about the detailed physical 
properties of disks. We also describe some additional forms of migration.
 \\~\\~\\~}
 
\end{list}
\end{abstract}

\section{INTRODUCTION}

Planet formation occurs in disks
of material that orbit young stars (see Chapter 13).
 At early stages of evolution, the disks
are largely gaseous
and have masses of typically a few
percent or more of the stellar masses. At later times,
after several $10^6 \y$,
the gas  disperses and disks largely consist of solid material.
Such disks last for timescales of  $\sim 10^8 \y$.
The interactions of a young  planet with its surrounding
disk  affects the planet's orbital energy and angular
momentum.
One consequence of such interactions is that
a planet may move radially through the disk (toward or away
from the central star), a process
called migration that is the topic of this chapter. 
In this chapter, we mainly concentrate on migration caused by gaseous disks.
Although gaseous disks survive only a small fraction of the lifetime
of the star and planet ($\sim 10^{10} \y$), 
calculations of disk-planet interactions described in this chapter
  show them to
be strong enough to have caused substantial migration.
In fact, a major problem is that the effects of migration
are typically predicted to be too strong to have allowed the
formation of gas giant planets to complete. That is, the typical
timescales for migration are found to be shorter than
both the lifetimes of gaseous disks and the predicted
formation timescales of gas giant planets in the so-called core accretion
model of $\sim 10^6 \y$ (see Chapter 13). Migration models applied some simple disk 
structures predict that
a forming planet should fall into (or very close to) the central star before it grows in mass to become a gas giant
({\em Ward,} 1997).

Early theoretical studies of planet migration, 
such as {\em Goldreich \& Tremaine}, 1980,
understandably concentrated on the
role of migration for the planets in our solar system. 
These studies suggested that there was substantial
migration of Jupiter while it was immersed in the solar
nebula. Migration should have occurred
both at Jupiter's early stages of formation
when it was a solid core  and at later stages when it was
a fully developed
gas giant planet interacting with the solar nebula.
The migration timescales were estimated to be much shorter
than lifetime of the nebula, although the direction of migration
was not determined.
However,
the evidence for migration was not  apparent.
The  location of Jupiter is just outside the snow-line,
where conditions for  rapid core formation are most favorable (see Chapter 13).
This situation seemed to suggest that Jupiter formed and remained
near its current location.
 Therefore, a plausible
conclusion was that migration did not play an important
role for Jupiter and therefore perhaps for all planets. 

Evidence for
the importance of migration changed dramatically in 1995 with the discovery
of the first giant planet, 51 Peg b, that has an
orbital period of only about 4 days  ({\em Mayor and Queloz}, 1995).
Subsequent discoveries of many giant planets with orbital
radii substantially smaller than Jupiter's 
added to the evidence (e.g., {\em Butler et al}, 2006).
Their existence suggested that
migration had occurred, since giant planet formation
close to a star is not likely to occur  ({\em Bodenheimer et al}, 2000;
{\em Ida \& Lin}, 2004).

 We might crudely think of a planet
orbiting in a gaseous disk around a star as being similar to a satellite in orbit about the
Earth. Atmospheric gas drag on the satellite causes it to lose
angular momentum and spiral down (migrate inwards) towards the Earth.
 However, the disk-planet situation is somewhat
different.  The planet and disk are both in orbit
about the star, and there is a
relatively small difference between the rotational speeds of the planet
and its neighboring gas. Although the velocity difference 
can produce an important level of gas drag for small
mass solid objects, gas drag does not produce the dominant torque
for objects as large as planets, as is discussed further in Section 2.1. For disk-planet interactions,
gravitational torques play the dominant role. 

The gravitational interactions between a planet and a gaseous disk are
difficult to analyze in a precise way. But a general picture has developed.
The results depend somewhat on the disk structure. We consider 
the disk to be what is called an accretion disk.
Accretion disks occur in many astronomical objects,
such as binary star systems and active galactic nuclei. Such disks
typically arise as gas flows towards a central body with some rotation about that body. 
Due to its angular momentum, the gas cannot
fall directly onto the central object and instead forms a disk in near
centrifugal balance. 
Observations of such systems, including disks around
young stars, reveal evidence that the gas is not simply orbiting about the
star, but also flowing inward (accreting) towards the central object.
The inflow velocity within the disk is very small compared to the orbital velocity.
But it is sufficient to reveal various observational signatures  as a consequence of this mass flow
towards and onto the central object. In particular, the inflow can result in the emission of radiation
from the accreting gas as it moves deeper
in the gravitational potential of the central object. In the case of interest to planet migration,
the gaseous disk orbits about a young central star. These so-called protoplanetary disks 
are described in Chapter 11.

Turbulence within
the disk is thought to be the main driver of the accretion.  The disk does not rotate about
the star at a constant rate with distance from the star. Instead, as a consequence of
centrifugal balance, it undergoes differential rotation, and the gas rotation
rate decreases with distance from the star. The gas orbits at (nearly)
the so-called Keplerian orbital frequency at orbital radius $r$ from the star of mass $M_{\rm s}$ given by
\begin{equation}
\Omega(r) = \sqrt{\frac{G M_{\rm{s}}}{r^3}}.
\label{omk}
\end{equation}
In a simple model, the disk turbulence can be considered to cause an effective friction.
Neighboring rings of gas in the disk interact by frictional forces as they rub against
each other due to the differential rotation. This friction
produces orbital energy losses and a torque that causes nearly all the material within the disk to move in (accrete), while angular
momentum  is transported outward ({\em Lynden-Bell \& Pringle,} 1974).
Instabilities within the disk are the cause of the turbulence.
The most likely candidates for causing the instabilities are 
disk self-gravity (e.g., {\em Paczynski}, 1978; {Lodato \& Rice}, 2004) 
and magnetic fields ({\em Balbus \& Hawley}, 1991). Disk self-gravity instabilities are more likely 
to be important at the earliest stages of disk evolution, while it is more massive.

\begin{figure}
 \epsscale{1.0}
 \plotone{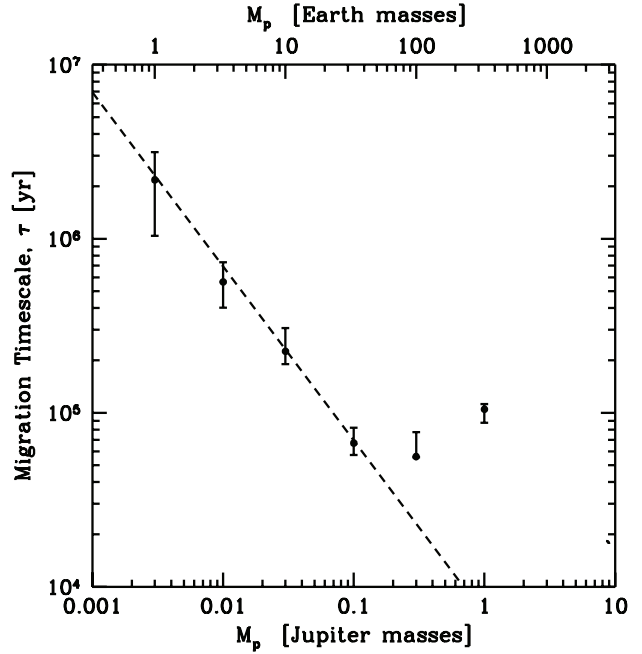}
  \caption{\small Migration timescales versus planet mass for a planet embedded in
  a 3D disk of mass 
  $\sim 0.02 M_{\rm s}$ with surface density $\Sigma \propto r^{-1/2}$  and $H/r = 0.05$.
  The planets are on fixed circular orbits and have fixed masses and orbit a solar mass star.
  The dots with error bars denote results of 3D numerical simulations
  with the same disk parameters
  ({\em Bate et al}, 2003).
  The  dashed line plots equation (\ref{tan}) based on linear theory ({\em Tanaka et al}, 2002). 
  Above about $0.1 M_{\rm J}$, the planet opens a gap in the disk, 
  Type I theory becomes invalid, and Type II migration occurs.} 
  \label{f:mig-time}
 \end{figure}

For small mass planets, the density structure of the disk is largely unaffected
by the presence of the planet. This situation leads to the so-called Type~I
regime of planet migration.
 In this simple model, the disk density structure is determined
by the disk turbulence. 
The disk differential rotation implies that gas that lies outside the orbit of the planet moves more slowly
 than gas inside the orbit of the planet. 
 The planet disturbs the disk
 by its gravity as gas passes near it. The disk disturbances act back on the planet and cause
 a net torque to be exerted on the planet.  This torque causes the planet to migrate.
 An analysis, such as in Section 2.2, shows that the migration
 timescale (orbit decay timescale) for a 10 Earth mass planet embedded in a disk that is somewhat similar
 to the minimum mass solar nebula ({\em Hayashi}, 1981) is only about $10^5 y$ (see Fig.~\ref{f:mig-time}).  Furthermore,
 the migration timescale varies inversely with planet mass. This result has the consequence that 
 planets migrate even faster as they grow in mass.  The shortness of
 this timescale poses a challenge to our understanding of migration,
 as discussed earlier. 
 
 For  a sufficiently high mass planet, its tidal torques can overpower the
 viscous torques associated with disk turbulence, and the planet
 can alter the disk density in its vicinity. 
 The result is that the planet
 opens a gap in the disk about its orbit. In this situation, we
 have the so-called Type~II regime of planet migration. The disk-planet
 interactions in this case are still gravitational. But the  gravitational
 torque on the planet is reduced because there is less material close
 to the planet compared to the Type~I regime. In the limiting
 case that the planet mass is small compared to the disk mass, 
 the planet moves inward like a test particle along
 with the accretion inflow of the gas. 
 The disk, being much more
 massive than the planet, sets the migration rate of the planet to its
 inflow rate. The disk density near the gap region adjusts so that the
 torque on the planet matches the disk's inflow speed. 
 
 The mass required for the planet
 to open a gap, as well as the migration rate, depends on the level of 
 disk turbulence. For typically adopted disk parameters, the transition
 to Type~II migration occurs at a few tenths of a Jupiter mass. The
 migration timescales are about $10^5 \y$. This timescale is also
 short compared to the disk lifetimes,
 but is much longer than it would have been if the planet had continued to
 undergo Type~I migration  (see Fig.~\ref{f:mig-time}). 
 In this idealized Type~II regime of high disk-to-planet
 mass ratio, the planet migration rate
 is independent of planet mass, since it is determined by the disk only.
  But for less extreme disk-to-planet mass ratios, the inertia of the planet
 plays a role and the migration rate decreases somewhat with planet mass.
 
 The determination of planet migration rates is not easy and various aspects remain 
 controversial. There are several reasons for this. 
 Although the interaction between a planet and its surrounding disk
 is gravitational, the interaction behaves in a way that is reminiscent
 of friction (dynamical friction). Gas that lies somewhat  exterior (interior) to the planet's
 orbit rotates slower (faster) than the planet and provides an inward (outward) torque on the planet
 as a consequence of the interaction. The net torque therefore
 involves a competition between nearly equal and opposite torques.
 Accurate calculations of the inward and outward torques are then required
 to determine the migration rate and the direction of migration.
   Another complication is that the disk-planet interactions generally get stronger
  closer to the planet. The results therefore depend on the details of
  how these torques become limited near the planet. 
  
\begin{figure}
 \epsscale{1.0}
 \plotone{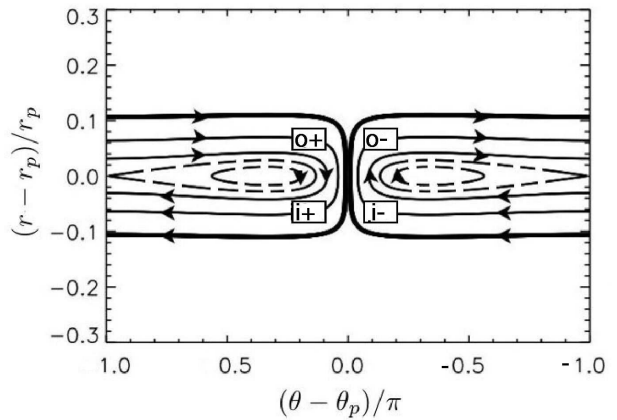}
  \caption{\small Coorbital streamlines near a Saturn mass planet that orbits a solar mass star,
  $M_{\rm p} = 3 \times 10^{-4} M_{\rm s}$. The origin is centered on the planet and corotates with it.
   Azimuthal angle $\theta$
  about the star
  increases to the left. The outermost streamlines are at the edge of the
  coorbital region. Solid streamlines
  are the horseshoe orbits. Locations $i\mp$ and $o\mp$ label
  the positions near the encounter with the planet.   }
  \label{f:co-traj}
 \end{figure}

  Another issue involves the effects
  of material that lies very close to the planet. This so-called coorbital region
  involves gas that does not circulate past the planet (see Fig.~\ref{f:co-traj}). Instead, the gas undergoes abrupt changes
  in its angular momentum due to encounters with the planet that take it
from interior of the planet's orbit radius to exterior of the planet's orbital radius and visa-versa.
The torques resulting from this region are quite different from those involving more weakly
perturbed material that lies
outside this region. The two regions, the weakly deflected and strongly deflected
regions, are dynamically quite different and require separate descriptions. Yet another issue
is that the torques depend on the detailed conditions within the disk. The disk structural variations,
i.e., how the disk density and temperature vary in radius,
can play an important role in  determining the torque. The detailed structural properties of disks
around young stars are not well understood and could be complicated (e.g., {\em Armitage et al}, 2000; 
{\em Terquem}, 2008). The resulting structures are likely 
more complicated than the simple power-law density distributions of the
minimum mass solar nebula model.  The level of disk turbulence
in a protostellar disk influences the disk structure and the rate of planet migration.
The amount turbulence   
 can vary with distance from the star and in time,
but it is not well understood.   
 
In Section 2 we describe calculations of   
migration rates due to disk-planet interactions. Section 2.1
discusses the effects of gas drag. Section 2.2
describes the migration rate of a low mass planet
interacting with a disk, modeled simply as a set of ballistic
particles that undergo mild or weak interactions with the planet.
This leads to an estimate of the migration rate, but
the approximations are not accurate enough to
 determine the direction of migration.
 Sections 2.3 discusses the torques due to
 material that comes very close to the planet, torques in the coorbital region.
 Section 2.4 discusses the more accurate determination of Type~I migration
 rates for a gaseous disk and the factors affecting migration.
Section 3 describes some of the outstanding issues:
how planet migration may be slower than estimated in Section 2,
some effects in planet migration that are missing from the idealized models
in Section 2, some forms of migration
not involving a gaseous disk, and some issues about the techniques used
in carrying out numerical simulations. 
Section 4 describes some future prospects.


\bigskip
\centerline{\textbf{ 2. Migration Rates Due To Disk-Planet Interactions}}

\bigskip
\noindent
\textbf{ 2.1 Aerodynamic Gas Drag}

\bigskip
Aerodynamic gas drag provides the dominant influence on the
orbital evolution of low mass solids embedded
in a gaseous disk that orbits a star ({\em Weidenschilling}, 1977;
{\em Cuzzi \& Weidenschilling}, 2006).  
The gas is subject to a radial
pressure force, in addition
to the gravitational force of the star. For a disk with smooth structural
variations in radius, this force induces slight departures
from Keplerian speeds of order $(H/r)^2 \Omega r$,
where $H$ is the disk thickness and $\Omega$ is the angular 
speed of the disk at radius $r$ from the star. 
For a thin disk ($H \ll r$) 
whose pressure declines in radius, as is typically
expected, the pressure force acts radially outward.
The gas rotation rate required to achieve centrifugal balance is then slightly below the Keplerian
rate. 
Sufficiently high mass solids are largely dynamically decoupled from the gas and
orbit at nearly the Keplerian rate, but are subject to
drag forces from the more slowly rotating gas. The drag leads
to their orbit decay. Both the 
drag force and the inertia increase with the size of the solid, $R$.
The drag force on the object increases with its area ($\sim R^2$), while its inertia
increases with its volume ($\sim R^3$). In the high mass solid regime, the dominance of inertia over drag
causes the orbital decay rate to decrease with object size as $1/R$ or with object mass as 
$1/M^{1/3}$.

The orbital migration rate due to gravitational
interactions between an object and a disk increases with the mass
of the object, as we will see in the next section. There is then a cross-over mass
above which the orbital changes due to disk gravitational forces
dominate over those due to drag forces.
For typical parameters, this value is much less than an Earth mass, $1 M_\oplus$
({\em Hourigan \& Ward}, 1984).
Consequently, for the purposes of planet migration, we will
ignore the effects of aerodynamic gas drag.

\bigskip
\noindent
\textbf{ 2.2 Torques Due To Mildly Perturbed Particles}
\bigskip

As an initial description of gravitational disk-planet interactions,
we consider the disk to consist of noninteracting particles, each having
mass $M$ much less than the mass of the planet $M_{\rm p}$.
The disk is assumed to extend smoothly across the
orbit of the planet without a gap. The planet is then in the Type~I regime of planet migration,
as discussed in the Introduction.
The model leads to insights about several issues concerning planet migration.
In particular, it explains why the migration rate increases with planet mass
and allows us to estimate the magnitude of the migration rate. This estimated migration rate
is numerically close to what more detailed calculations reveal for simple disk structures. The approximations
fall short of allowing a determination of whether the migration is inward
or outward and the determination of the detailed dependence of the migration rate on disk properties, such
as the density and temperature distributions. The treatment describes material that
orbits about the star and undergoes a small deflection by the planet. It does not describe the effects
of material that lies very close to the planet in the so-called coorbital region, where the deflections are large (see Fig.~\ref{f:co-traj}). We
will consider a model for that region in Section 2.3.

In this model, the particles are considered to encounter and pass by the vicinity of a planet
that is on a fixed circular orbit of radius $\rp$ about the star of mass 
$M_{s}$. 
As a result of their interactions, the planet and disk exchange
energy and angular momentum. 
This situation
is a special case of the famous three-body problem
in celestial mechanics.
The tidal or Hill radius of the 
planet where planetary gravitational forces dominate
over stellar and centrifugal  forces is given by
\begin{equation}
R_{\rm H} = \rp \left(\frac{M_{\rm p}}{3 M_{\rm s}} \right)^{1/3},
\label{Hill}
\end{equation}
where radius $R_{\rm H}$ is measured from the center of the planet.

\begin{figure}
 \epsscale{0.6}
 \plotone{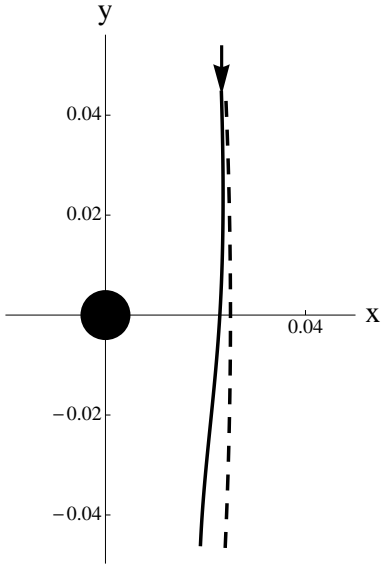}
 \caption{\small Path of a particle that passes by a planet
 of mass $M_{\rm p} =10^{-6} M_{\rm s}$ (0.3 Earth masses for a planet that orbits
 a solar mass star). The coordinates are in units
 of the orbital radius of the planet $\rp$.
 The planet lies at the origin, while the star lies at $(-1,0)$.
 The dashed line follows the
 path that is undisturbed by the planet with $x=0.025 \rp$, while the solid
 line follows the path resulting from the interaction with the planet.
 In the frame of the planet, the particle moves in the negative $y$
 direction.
 }
 \label{f:traj}
 \end{figure}

Consider a particle
that approaches the planet on a
circular orbit about the star with orbital radius $r$ sufficiently different from
$\rp$ to allow it to freely pass by the planet with a small
deflection. This condition requires that the closest approach
between the particle and planet $\sim |r-\rp|$ 
to be somewhat greater than a few times $R_{\rm H}$.
The solid line in Fig.~\ref{f:traj} shows the path of a particle  deflected by a planet
 whose mass
is $10^{-6} M_{s}$ (corresponding to 0.3 Earth masses for a planet that orbits 
a solar mass star), Hill radius $R_{\rm H} \simeq 0.007 \rp $, and
orbital separation $r - \rp \simeq 3.5 R_{\rm H}$. The planet and particle orbit counter-clockwise in the inertial frame
with angular speeds $\Omega_{\rm p}$ and $\Omega(r)$, respectively.
In the frame of the planet, the particle moves downward
in the figure, since its angular speed is slower than that
of the planet ($\Omega(r) < \Omega_{\rm p}$ for $r>\rp$).

To estimate the angular momentum change of a
particle like that in Fig.~\ref{f:traj}, we consider a Cartesian coordinate
system centered on the planet and corotates with it.
The $x$ axis lies
along a line between the star and planet and points away from the star. 
We consider in detail the fate of a particle that starts its approach toward
the planet on a circular orbit of orbital radius $r$ that lies outside the orbit of the planet, $
x = r-\rp>0$.
The planet and the pre-encounter particle are on circular Keplerian orbits whose
orbital frequencies are given by equation (\ref{omk}).
We later discuss what happens to particles that begin their encounter
at orbital radii smaller than the planet's ($x<0$).
The dashed
line in Fig.~\ref{f:traj} traces the path 
that the particle would take in the absence of the planet,
while the solid line shows the path in the presence of the planet.
In both cases, the particle paths are generally along  the negative $y$ direction.
The two paths are nearly identical 
prior to the encounter with the planet (for $y>0$).

The  particle velocity in the frame that corotates with the planet is then approximately
given by 
\begin{equation}
{\bf v}  \simeq r (\Omega(r) - \Omega_{\rm p}) {\bf e_y} \simeq x \, \rp \frac{d \Omega}{dr} {\bf e_y}
\simeq -\frac{3}{2} \Omega_{\rm p} 
x \, {\bf e_y},
\label{up}
\end{equation}
where $x= r - \rp$, $\Omega(r)$ is given by equation (\ref{omk}), and $\Omega_{\rm p}=\Omega(\rp)$.
Upon interaction with the planet, the particle is deflected
slightly toward it. The particle of mass $M$ experiences a force
$F_x = -G M M_{\rm p} x /(x^2 + y(t)^2)^{3/2}$. This force mainly acts over a time $t$
when $|y(t)| \la x$ and is $F_x \sim -G M M_{\rm p} /x^2$. 
From equation (\ref{up}), it follows that
the encounter time 
$\Delta t \sim x/v \sim 1/\Omega_{\rm p}$,
of order the orbital period of the planet, independent of $x$. 

To proceed, we apply the so-called impulse
approximation, as employed by  {\em Lin \& Papaloizou}, 1979. The approximation involves the assumption
that the duration of the interaction is much shorter than the orbital
period. Since the duration of the encounter
is of order the orbital period, the approximation
is only marginally satisfied. Consequently, the 
expressions we obtain 
cannot be determined with high accuracy in this approximation.
They contain the proper dependences
on various physical quantities. But the approximation does not lead to 
 the correct dimensionless numerical coefficients of proportionality for
 the angular momentum change of a particle. Therefore, we suppress the
 numerical coefficients in the analysis.
An exact treatment in the limit of weak
perturbations exerted by a small mass planet is given in {\em Goldreich \& Tremaine},
1982. 

As a result of the
encounter, the particle with $x>0$ acquires an $x$ (radial) velocity 
\begin{equation}
v_x \sim  \frac{F_x \Delta t}{M} \sim - \frac{G M_{\rm p}}{x^2 \Omega_{\rm p}}.
\label{ux}
\end{equation}
The particle is then deflected by an angle 
\begin{equation}
\delta \sim \frac{v_x}{v} \sim \left (\frac{M_{\rm p}}{M_{\rm s}} \right) \left(
\frac{\rp}{x} \right)^3 \sim  \left(\frac{R_{\rm H}}{x} \right)^3 
\label{delta}
\end{equation}
after the encounter (see Fig.~\ref{f:traj}).

To determine the change in angular momentum of the particle, 
we need to determine its change in velocity  along
the  $\theta$ direction, that is the same as 
the $y$ direction near the planet.
To determine this velocity change, $\Delta v_y$, we ignore the effects of the star during the encounter,
and apply conservation of kinetic energy between the start and end of the encounter
in the frame of the planet. The velocity magnitude
$v$ is then the same before and after the encounter, although the direction
changes by angle $\delta$.  Since the particle in Fig.~\ref{f:traj} moves in the negative $y$ direction, its pre-encounter $y$ velocity
is $-v$ and its post-encounter $y$ velocity is $-v \cos{\delta}$.
We  then have that the  change of the 
velocity of the particle along the $y$ direction is
\begin{equation}
\Delta v_y =  - v \cos{\delta} + v,
\label{kec}
\end{equation}
where  $v \simeq 1.5 \Omega_{\rm p} 
x$ is the magnitude of the velocity before the encounter (see equation (\ref{up})).
We assume that the perturbation is weak, $\delta<< 1$, and obtain
from equations (\ref{up}), (\ref{delta}), and (\ref{kec})  
\begin{equation}
\Delta v_y \sim v \, \delta^2 \sim  \rp \Omega_{\rm p}  \left( \frac{M_{\rm p}}{M_{\rm s}} \right)^2 \left( \frac{\rp}{x} \right)^5.
\end{equation}
It then follows that the change in angular momentum of the particle
is given by
\begin{equation}
\Delta J \sim M \rp \Delta v_y \sim M  \rp^2 \Omega_{\rm p}  \left( \frac{M_{\rm p}}{M_{\rm s}} \right)^2 \left( \frac{\rp}{x} \right)^5.
\label{DJa}
\end{equation}

\begin{figure}
 \epsscale{0.8}
 \plotone{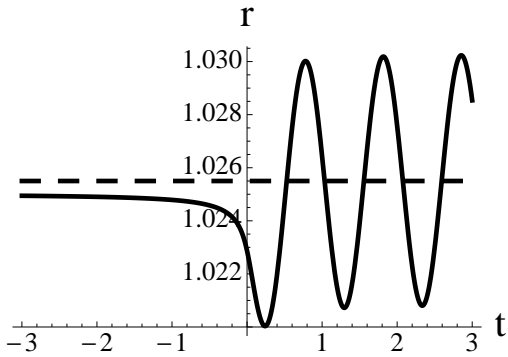}
 \caption{\small  The solid line plots the particle distance from the star
 $r$ in units of $\rp$  as a function of time in units of planet
 orbit periods  for the particle that follows the perturbed path in Fig.~\ref{f:traj}. The particle passes the planet at time $t=0$. Immediately after passage by the planet,
 the particle is deflected toward smaller radii, toward the planet, and acquires an eccentricity, as indicated by the radial oscillations. The dashed
 line plots the mean
 radius of these oscillations. Since the mean radius of the oscillations 
 is larger than the initial orbital radius (compare dashed line with solid line at $t<-1$), 
  the particle gained  energy and angular momentum,
 as a consequence of its interaction with the planet.
 }
 \label{f:r-t}
 \end{figure}

Fig.~\ref{f:r-t} plots the orbital radius as a function of time for the particle
plotted in Fig.~\ref{f:traj}, where $t=0$ is the time
of closest approach to the planet. Notice that after the encounter, 
the particle orbit acquires an eccentricity, as
seen by the radial oscillations, and an increased angular
momentum, as seen by the mean shift of the radius for the oscillations.
Although the particle is initially deflected inward (negative $x$ direction, towards the star),
as expected by equation (\ref{ux}),
it rebounds after the encounter to  an increased
time-average radius and higher angular momentum, as expected by equation (\ref{DJa}).

\begin{figure}
 \epsscale{1.5}
 \plottwo{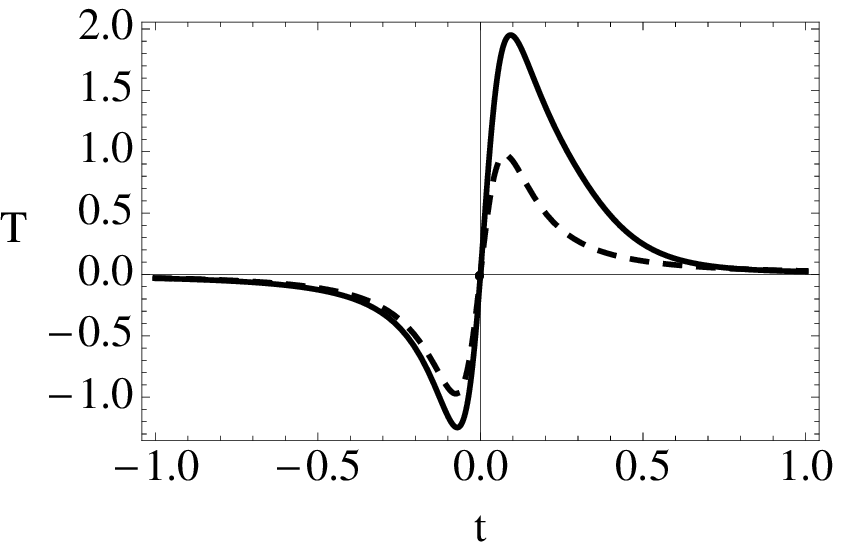}{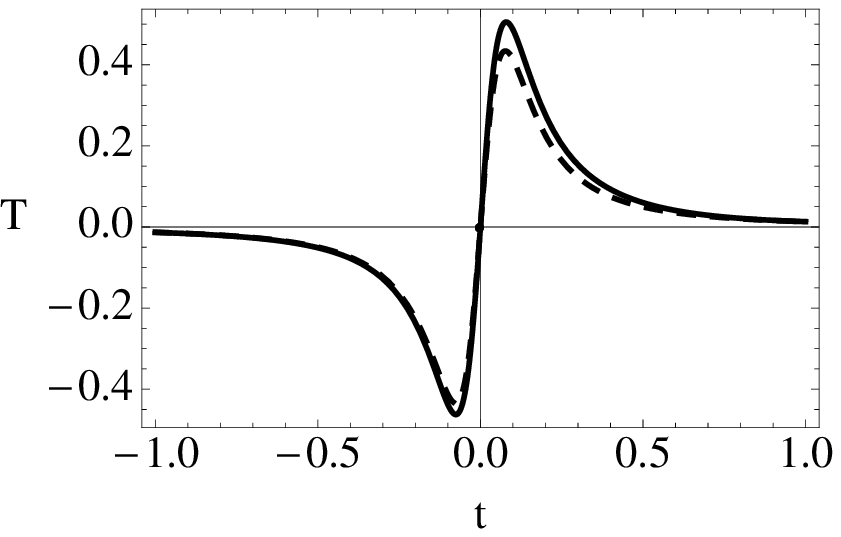}
 \caption{\small Torque on the particle of mass $M$ that is normalized by $10^3 M \Omega_{\rm p}^2 \rp^2$ as a function of time in units of the planet's
 orbit period along the unperturbed (dashed lines) and perturbed
 paths (solid lines). The planet mass $M_{\rm p}=10^{-6} M_{\rm s}$ (0.3 Earth masses for a planet that
 orbits a solar mass star).
 Top panel is for a particle having $x=0.02 \rp$, 
 the case in Fig.~\ref{f:traj}. The bottom panel is  
 for a particle with $x = 0.03 \rp$. 
 }
 \label{f:T-ta}
 \end{figure}

Fig.~\ref{f:T-ta} plots the torque on a particle due to the planet  as a 
function of time, where $t=0$ is the time of closest
approach between the particle and the planet. The solid curves are for particles that follow
their actual paths (similar to the solid line in Fig.~\ref{f:traj}).
The dashed curves are for particles that are made to follow
the unperturbed paths (similar to the dashed line in Fig.~\ref{f:traj}).
The net angular momentum change is the time-integrated
torque.
The torque on a particle 
along an unperturbed path  (such as the
dashed curve in Fig.~\ref{f:traj}) is antisymmetric
in the time $t$. Consequently, there
is no net change in angular momentum accumulated along
 this path. The change in angular momentum along the
 unperturbed path from $t = -\infty$  to $t=0$ 
 is linear in the planet mass, since
 the force due to the planet is proportional to its mass.
 The departures from antisymmetry of the torque versus time result in the 
net angular momentum change. These departures are a consequence
of the path deflection (solid curve in Fig.~\ref{f:traj}).

\begin{figure}
 \epsscale{2.0}
 \plottwo{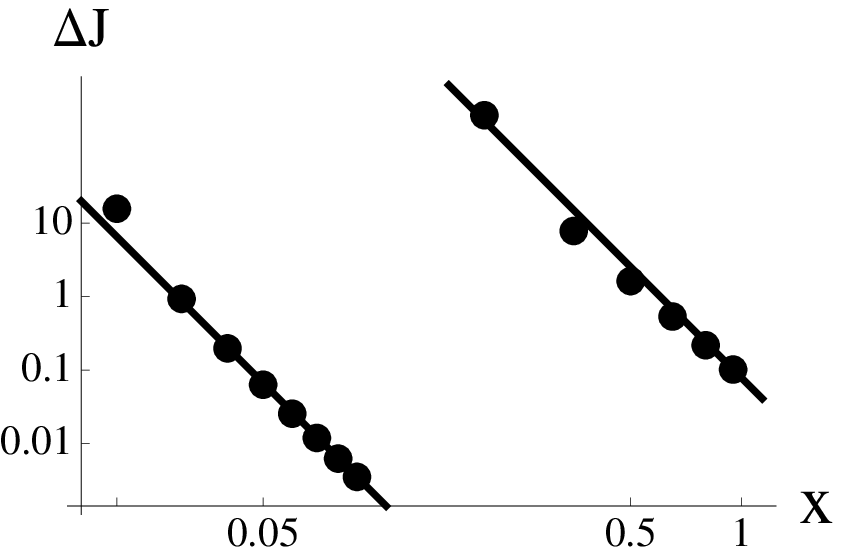}{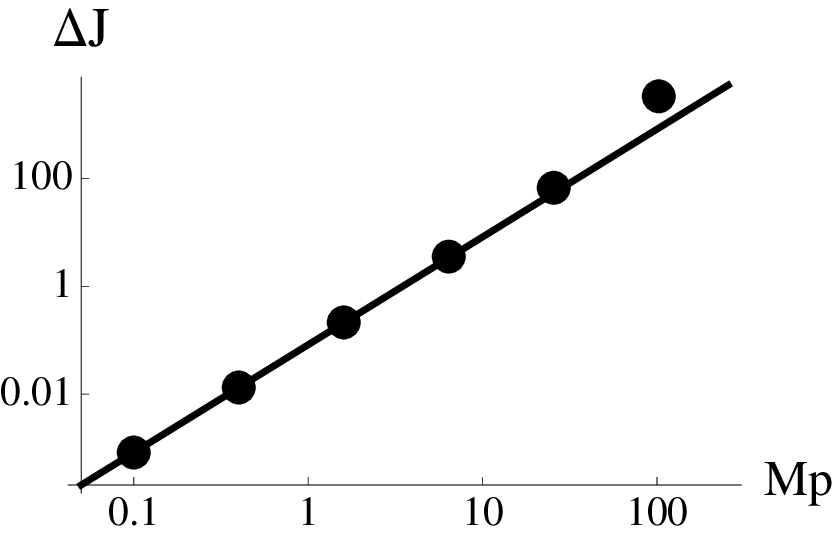}
  \caption{\small Numerical test of equation (\ref{DJa})
  based on orbit integrations. Top panel: Log-log plot
  of $10^{4} \Delta J/ (M  \rp^2 \Omega_{\rm p})$
  as a function of preimpact distance from the planet's orbit $x$ in units of $\rp$. The lower set of points is for
  a planet of mass $M_{\rm p} = 10^{-6} M_{\rm s}$ (0.3 Earth masses for a planet that
 orbits a solar mass star). The upper set is for
    $M_{\rm p} = 10^{-3} M_{\rm s}$ (a Jupiter mass for a planet that
 orbits a solar mass star). The solid lines are for $\Delta J \propto x^{-5}$ that pass through the respective right-most
    points. Bottom panel:  Log-log plot
  of $10^{5} \Delta J/ (M  \rp^2 \Omega_{\rm p})$ as a function of planet mass in Earth masses for a planet that orbits a solar mass star.
  The points are the results of numerical simulations for
  a fixed value of $x = 0.2 \rp$. The solid line is for  $\Delta J \propto (M_{\rm p}/M_{\rm s})^2$ that passes through the left-most point.
}
\label{f:djt}
 \end{figure}

For the case of the closer encounter plotted in the top panel of Fig.~\ref{f:djt}, the departures
from the unperturbed case are substantial. But for a
slightly larger orbit (bottom panel), 
 the torque along the perturbed path differs only 
slightly from the torque along the unperturbed path that integrates
to zero. This behavior is a consequence of the steep decline of $\Delta J$
with $x$ in equation (\ref{DJa}). For $x$ greater than a few $R_{\rm H}$,
the angular momentum
change acquired before the encounter is nearly equal and opposite to
the angular momentum change at later times $t>0$. The net
angular momentum change is given by equation (\ref{DJa}) and
is quadratic in the planet mass. The quadratic dependence on $M_{\rm p}$ is a 
consequence of the deviations in torque between the perturbed
and the unperturbed paths, since the angular
momentum change along the unperturbed path
is zero. These deviations involve the
product of the linear dependence of the force
on planet mass with the linear dependence of the path deflection
on planet mass (equation (\ref{delta})) and so are quadratic in $M_{\rm p}$.
This quadratic dependence of torque on planet mass is important.
It leads to the conclusion we obtain later that the migration rate
of a planet is proportional to its mass. 

The particle in Fig.~\ref{f:traj} gains angular momentum, as predicted by 
equation  (\ref{DJa}).
The reason is that 
the path deflection  occurs mainly after the particle
passes the planet, that is for $y<0$. The deflection takes the particle closer to the planet than
would be the case along the unperturbed path.
The planet then pulls
the particle toward positive $y$, causing it to gain angular 
momentum. Just the opposite would happen for a particle
with $r < \rp$, for $x<0$ in equation (\ref{DJa}). The particle
would approach the planet in the positive $y$ direction,
be deflected towards the planet for $y >0$, and be pulled
by the planet in the negative $y$ direction, causing it to lose angular
momentum. The interaction behaves somewhat like friction.
The particle gains (loses) angular momentum if it moves 
slower (faster) than
the planet. The angular momentum then flows outward as a
result of the interactions. That is, for a decreasing angular velocity with radius
($d \Omega /dr <0$) as in the Keplerian 
case,
a particle whose orbit
lies interior to the planet gives angular momentum
to the planet, since the planet has a lower angular speed than the particle. 
The planet in turn gives angular momentum to a particle
whose orbit lies exterior to it.

The planet in effect pushes material away from its orbit. A particle outside the
orbit of the planet is forced outward (in a time-averaged sense) as it gains angular momentum, and
a particle inside is forced inward as it loses angular momentum. From this point of view,
the gravitational effects of the planet behave in a repulsive manner.

Fig.~\ref{f:djt} shows the results of numerical tests of equation (\ref{DJa}). It verifies
the dependence of $\Delta J$ on $x$ and $M_{\rm p}$.
Departures of the expected dependences (solid lines)
occur when $x \simeq 3 R_{\rm H}$. At somewhat smaller 
values of $x$, the particle orbits
do not pass smoothly by the planet. Instead, they lie within the coorbital
region where they periodically undergo strong deflections, as seen in Fig.~\ref{f:co-traj}. We omit
this region from current consideration and consider it later in Section 2.3.

We now  determine the torque on the planet
for a set of particles that form a continuous disk (of zero thickness) that lies in the orbit plane
of the planet with surface density (mass per unit area)
$\Sigma$ that we take to be constant in the region near the planet.
The particle disk provides a flux of mass (defined to be positive) past the planet between $x$ and  $x+dx$
\begin{equation}
d\dot{M} \sim |v_y| \, \Sigma \, d x
 \sim \Sigma \, \Omega_{\rm p} \, x \, dx.
\end{equation}
We evaluate 
the torque $T_{\rm{out}}$ on the planet due to disk material
that extends outside the orbit of the planet from $r= \rp+ \Delta r$ to
 $\infty$ or $x$ from $\Delta r$ to  $\infty$, where $\Delta r > 0$. 
 We use the fact that the
 torque  the planet exerts on the disk is equal
 and opposite to the torque the disk exerts on the planet.
 We then have
\begin{eqnarray}
T_{\rm{out}} &\sim& -\int_{\Delta r}^{\infty} \frac{\Delta J}{M} \frac{d\dot{M}}{dx} dx, \\
    T_{\rm{out}}  & = & - C_{\rm T} \Sigma \rp^4  \Omega_{\rm p}^2 \left( \frac{M_{\rm p}}{M_{\rm s}} \right)^2 
        \left( \frac{\rp}{\Delta r} \right)^3,
\label{To}
\end{eqnarray}
where $C_{\rm T}$ is a dimensionless positive constant of order unity and $\Delta J$ is evaluated
through equation (\ref{DJa}).
The torque on the planet due to the disk interior to the orbit of the planet
from $x = -\infty$ to $x = \Delta r$ with $\Delta r<0$ evaluates to
$T_{\rm{in}}= - T_{\rm{out}}$ or
\begin{equation}
T_{\rm{in}}= C_{\rm T} \Sigma \Omega_{\rm p}^2 \rp^4 \left( \frac{M_{\rm p}}{M_{\rm s}} \right)^2 
        \left( \frac{\rp}{|\Delta r|} \right)^3.
\label{Ti}
\end{equation}

The equations of motion for particles subject to only  gravitational forces
are time-reversible.  We saw that particle in Fig.~\ref{f:traj} gains angular
momentum from its interaction with the planet.
But, if we time-reverse the particle-planet encounter
in Fig.~\ref{f:traj}, we see that the  eccentric orbit particle
would approach the planet (both on clockwise orbits) and
then lose angular momentum (apply $t \rightarrow -t$ in Fig.~\ref{f:r-t}).
What determines whether a disk particle gains or loses angular momentum?
We have assumed that the particles always approach
the perturber on circular orbits. 
The particles periodically encounter the gravitational effects of the planet.
The closer the particle orbits to the planet, the smaller the relative
orbital speeds and the longer the time
between encounters. From equation (\ref{up}), it follows that
the time between encounters $\tau \simeq 2 \pi \rp /v$
is estimated as 
\begin{equation}
\tau \sim \frac{\rp \,P}{|x|},
\label{tau}
\end{equation}
where $P$ is the planet orbital period. Since the
encounters are close $ |x| \ll \rp$, the time between encounters
is long compared to the encounter time $\sim P$, that is $\tau \gg P$.
But we saw
in Fig.~\ref{f:r-t} that the particles acquire eccentricity
after the encounter. For this model to be physically consistent,
we require that this eccentricity damp between encounters with the planet.
The eccentricity damping produces an arrow of time for the angular momentum
exchange process that favors 
circular orbits ahead of the encounter as shown in Fig.~\ref{f:traj}, resulting
in a gain of angular momentum for particles that lie outside the orbit of the planet.

Equations (\ref{To}) and (\ref{Ti}) have important consequences.
The torques on the planet arising from the inner and outer
disks are quite powerful and oppose each other.
This does not mean that the net torque on the planet is zero
because we have assumed perfect symmetry across $r=\rp$.
The symmetry is broken by higher order considerations.
For this reason, migration torques
are often referred to as differential torques.

Since the torques are singular in $\Delta r$, 
they are dominated by material that
comes close to the planet. Consequently,
the asymmetries occur through differences
in physical quantities
at radial distances $\Delta r$ from the planet.
In a gaseous disk, the effects of both temperature and density variations in radius
can play a role in the asymmetry, as well as asymmetries associated with
the differential rotation of the disk.  We cannot describe the asymmetries due
to temperature in this model because temperature is not described by the ballistic
particles.
For this simple model of a (pressure-free) particle disk,
we consider for example the effect of the density
variation in radius that we have ignored up
to this point, although the other asymmetries are important. It does
not matter which quantity is considered for the purposes of obtaining
a rough expression for the net torque. 

We expand the disk density in a 
Taylor expansion about the orbit of the planet
and obtain that
\begin{equation} 
\Sigma_{\rm{out}} - \Sigma_{\rm{in}} \simeq
2 \Delta r \frac{d\Sigma}{dr},
\label{sigas}
\end{equation} 
where 
$\Sigma_{\rm{in}}$ and $\Sigma_{\rm{out}}$
are the surface densities at $r=\rp-\Delta r$
and $r=\rp+\Delta r$, respectively. 
Consequently, for $ |d\Sigma/dr| \sim \Sigma_{\rm p}/\rp$,
we expect that the sum of the inner and outer torques
to be smaller than their individual values by an amount of
order $\Delta r/\rp$. Similar considerations apply to variations
in other quantities.
That is, we have that the absolute value of the net torque $T$ on the
planet is approximately given by
\begin{eqnarray}
|T| &=& |T_{\rm{in}} + T_{\rm{out}}|, \\
     &\sim&  |T_{\rm{in}}| \, \frac{|\Delta r|}{\rp}, \\
     &\sim& \Sigma \, \Omega_{\rm p}^2  \, \rp^4 \, \left( \frac{M_{\rm p}}{M_{\rm s}} \right)^2 
        \left( \frac{\rp}{|\Delta r|} \right)^2.
        \label{Tnet}
\end{eqnarray}

The above equation for the torque (equation (\ref{Tnet}))
must break down for small $\Delta r$,
in order to yield a finite result. For  values of $|\Delta r| \la 3 R_{\rm H}$, the particles become 
trapped in closed orbits in the so-called coorbital region
(see Fig.~\ref{f:co-traj}). We exclude this region
from current considerations, since the torque derivation we considered
here does not apply in this region. In particular,
the assumption that particles pass by the planet with
a small deflection is invalid in this region.
Using equation (\ref{Tnet}) with $|\Delta r| \sim R_{\rm H}$, we obtain a   
torque
\begin{equation}
|T|  \sim \Sigma \, \Omega_{\rm p}^2 \, \rp^4  \left( \frac{M_{\rm p}}{M_{\rm s}} \right)^{4/3}.
\label{Tnp}
\end{equation}

Another limit on $\Delta r$ comes about due to  gas
pressure.  One effect of gas pressure
is to cause the disk to have a nonzero thickness $H$.
The disk thickness is measured by its scale height out of the orbital plane.
The disk scale height is determined by the force balance 
in the direction perpendicular to the orbit plane, the $z$ or vertical direction. Gas pressure forces act to
spread the gas in this direction, while gravitational forces
act to confine it. We consider the vertical structure
of an axisymmetric disk, unperturbed by a planet.
 There are no motions in the vertical direction
and the disk is said to be in hydrostatic balance. 

The vertical hydrostatic balance condition can be written as
\begin{equation}
\frac{ \partial p(r,z)}{\partial z} = \rho g_z,
\label{heq}
\end{equation}
where $p(r,z)$ is the gas pressure, $\rho(r,z)$ is the gas density, and $g_z(r,z)$ is the vertical
gravity.
The vertical gravity is the $z$ component of the gravitational force per unit disk mass due to the star
that is equal to 
\begin{equation}
g_z(r,z) =-\frac{G M_{\rm s} z}{r^3} = -\Omega^2(r)\, z,
\label{gz}
\end{equation}
where $\Omega$ is given by equation (\ref{omk}).
For a vertically isothermal disk, a disk whose temperature and therefore sound speed $c(r)$
is independent of $z$, we have
\begin{equation}
p(r,z) = \rho(r,z) \, c^2(r).
\label{p}
\end{equation}
Substituting equations (\ref{gz}) and (\ref{p}) into equation
(\ref{heq}), we obtain
\begin{equation}
\rho(r,z) = \rho(r,0) \exp{\left( -\frac{z^2}{2 H^2} \right)}
\label{rhoz}
\end{equation}
and
\begin{equation}
H(r) = \frac{c(r)}{\Omega(r)},
\label{H}
\end{equation}
where $c$ is the gas sound speed that is a function
of disk temperature. Disk thickness $H$
is a measure of the importance of gas pressure.
For conditions in protostellar disks, the disk thickness to radius
ratio (disk aspect ratio) is typically $0.03 \la H/r \la 0.1$,
corresponding to a gas sound speed of about $1.5 \, \kms$ at $1 \AU$
from a solar mass star.

For an axisymmetric disk of nonzero thickness, the surface density $\Sigma = \int_{-\infty}^{\infty} \rho \, dz$
evaluates to
\begin{equation}
\Sigma(r) = \int_{-\infty}^{\infty} \rho(r,z) \, dz = \sqrt{2 \pi} H(r) \rho(r,0),
\end{equation}
where $\rho(r,z)$ is defined by equation (\ref{rhoz}).
The gas density is in effect smeared over distance 
$H$ out of the orbit plane. Near the planet, the  gas gravitational
effects are then smoothed over distance $H$. Distance
$\Delta r$
is in effect limited to be no smaller than $\sim H$.  For $\Delta r \sim H$, the torque 
expression (\ref{Tnet}) is then
estimated as
\begin{equation}
 |T| \sim \Sigma \, \Omega_{\rm p}^2 \, \rp^4 \left( \frac{M_{\rm p}}{M_{\rm s}} \right)^2 
        \left( \frac{\rp}{H} \right)^2.
 \label{Tp0}
\end{equation}

Which form of the torque applies (equation (\ref{Tnp}) or (\ref{Tp0}))
to a particular system
depends on the importance of gas pressure.
For $H < R_{\rm H}$, gas pressure effects are small compared
to the gravitational effects of the planet near the Hill radius.
Consequently, we expect
equation (\ref{Tnp}) to be applicable in the case  of relatively
weak gas pressure, $H < R_{\rm H}$,
and equation (\ref{Tp0}) to be applicable otherwise.
For typically expected
conditions in gaseous protostellar disks, it turns out
that equation (\ref{Tp0})
is the relevant one for planets undergoing Type I
migration.

The migration rate $\dot{r}_{\rm p}/\rp = T/J_{\rm p}$ for a  planet with angular 
momentum $J_{\rm p}$ is then
linear in planet mass, since $T$ is quadratic
while $J_{\rm p}$ is linear in planet mass.
Therefore, the Type I migration rate
increases with planet mass, as asserted in Section 2.1.
This somewhat surprising result that more massive
planets migrate faster is in turn a consequence
of the quadratic variation of $\Delta J$ with planet mass
in equation (\ref{DJa}). This quadratic dependence occurs because the
possible linear dependence of $\Delta J$ on planet mass
vanishes due to the antisymmetry of the torque
as a function of time along the unperturbed particle
path, as discussed after 
equation (\ref{DJa}).

 Based on equation (\ref{Tp0}) with typical
 parameters for the minimum mass solar nebula at the location of Jupiter
  $\rp= 5 \AU$ ($\Sigma =150 \, \sdunits$, 
 $\Omega=1.8\times 10^{-8} \, \si$, and
 $H=0.05 \rp$), we  estimate the planetary migration
timescale $J_{\rm p}/T$ for a planetary core of 10 Earth masses  embedded 
in a minimum mass solar nebula as $4 \times 10^5 \y$.
This timescale is short compared to the disk lifetime, estimated as 
several  times $10^6 \,\y$, or the Jupiter formation timescale of $\ga 10^6 \, \y$ in
the core accretion model. 
The relative shortness of the migration
 timescale is a major issue for understanding
planet formation. Since migration is found
to be inward for simple disk models, as we will see later, the timescale disparity
suggests that a planetary core will fall into the central star before
it develops into a gas giant planet.
Research on planet migration has concentrated
on including additional effects such as gas pressure
and on improving the migration rates by means
of both analytic theory and multi-dimensional
simulations.

A more detailed analysis reveals that the 
torque in equation (\ref{Tp0}) does provide
a reasonable estimate for the magnitude of migration rates
in gaseous disks in the so-called Type I regime
in which a planet does not open a gap in the disk.
However, the derivation of the torque in this section is not precise enough
to determine  
whether the migration is inward or outward (i.e.,
whether $T$ is negative or positive). The density asymmetry
about $r=\rp$ (see equation (\ref{sigas}))
  typically 
involves a higher density at smaller $r$, as in the case of the
 minimum mass solar nebula. This variation  suggests that torques
from the inner disk dominate, implying outward migration, as was thought
to be the case in early studies. But this conclusion is incorrect.
As we will discuss later, inward migration is typically favored, at least for simple 
disk models. 
We  have not included  the effects of gas temperature and gas pressure. 
A disk with gas pressure  propagates density waves launched by the planet. 
The analysis in this section has only considered effects of material that passes
by the planet. In addition, there are effects from material that
lies closer to the orbit of the planet (see Fig.~\ref{f:co-traj}). 
This region can also provide torques.
We have also assumed that the disk density is undisturbed by 
presence of the planet. Feedback effects of the disk disturbances
and gaps in the disk can have an important influence on migration.
Finally, there are other physical effects such as disk turbulence
that should be considered. We will consider such effects in subsequent sections.

\bigskip
\noindent
\textbf{ 2.3 Coorbital  Torques}
\bigskip

Thus far we have considered torques that arise
from gas that passes by the planet in the azimuthal direction
(the $y$ direction in Fig.~\ref{f:traj}) with a modest deflection. 
Gas  that resides closer to the planet, $ | r- \rp| \la 3 R_{\rm H}$,
in the so-called coorbital region,
does not pass by the planet. 
Instead, it follows what are called librating
orbits in the corotating frame
of the planet as seen in Fig.~\ref{f:co-traj}. We will consider librating orbits of particles
that reach close to a planet that lies on a fixed circular orbit, the so-called horseshoe orbits.
A particle at position $o+$ in Fig.~\ref{f:co-traj}
is in circular motion outside the orbit of the planet. 
It  moves more slowly than the planet
and approaches it. The particle in this case is more strongly
perturbed by the planet than the more distant particle in Fig.~\ref{f:traj}.
It gets pulled inward by the planet, causing it to change to a circular
orbit interior to the planet's orbit once it reaches position $i+$. Since the particle is now moving faster than
the planet, it moves away from it. In the process, the particle has executed a U-turn. 
At a later time, the particle approaches the planet from behind, at position $i-$, and the planet
pulls the particle outward, causing it to make a second U-turn to position $o-$.
The particle is then back to the initial outer radius that it started on before both encounters.
The particle path is closed in the corotating frame of the planet.

With each change in angular momentum of the particle, there is an equal and opposite change
in angular momentum of the planet.
But, whatever angular momentum is gained by a particle as
it changes from the inner radius $r_{\rm{i}}$ to outer
radius $r_{\rm{o}}$  is lost when it later
encounters the planet and shifts from $r_{\rm{o}}$ to $r_{\rm{i}}$.
The reason is that
the particle follows a periodic orbit in the frame of the planet, so there
is no change in angular momentum over a complete
period of its motion.
As a result, angular momentum changes do not grow over long
timescales. At any instant, the largest angular momentum change possible over any
time interval
is that acquired in the last particle-planet encounter. 
 Therefore, the time-averaged torque on the planet
due to a particle drops to zero over times scales longer than the period of its motion.
So although  particles on closed horseshoe orbits can come close to the planet,
the orbital symmetry  limits the torque that
they exert on the planet.  

\begin{figure}
 \epsscale{0.75}
 \plotone{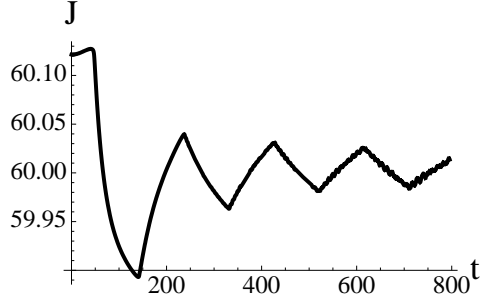}
  \caption{\small 
  Total angular momentum for a set of 60 particles, each of mass $M$,
  on horseshoe orbits in units of $M \rp^2 \Omega_{\rm p}$
  as a function of time in units of the planet orbit period in a star-planet system with 
  $M_{\rm p} = 3 \times 10^{-6} M_{\rm s}$ (0.3 Earth masses for a planet that orbits a solar mass star). 
  The particles start at $t=0$ distributed
  between $r= \rp+ R_{\rm H}/60$ to $r= \rp+ R_{\rm H}$ 
  (with $R_{\rm H} \simeq 7\times10^{-3} \rp$) along the
  star-planet axis 180 degrees from the planet. The changes in angular
  momentum cause a torque to be exerted on the planet. The torque (the time derivative of $J$)
  oscillates  and declines in time as angular momentum becomes more constant in time because the particles undergo phase mixing on the libration
  timescale $\sim 150$ planet periods.
   }
   \label{f:Jco}
 \end{figure}

For a set of particles, there is a further type of  torque cancellation that occurs.
Recall that the time period between particle-planet 
encounters (the libration timescale) varies inversely with the particle-planet orbital separation as $1/x$ (see equation (\ref{tau})).
So even if  a set of particles at different radii are  initially lined up  to encounter the planet at the same
time and produce a large torque, they will eventually drift apart in azimuth (phase) 
and encounter the planet at different times.  
After a while, at a time when
one particle gains angular momentum another will be losing it.  This randomization,
called "phase mixing", leads to a drop in torque over time.
Fig.~\ref{f:Jco} shows the angular momentum evolution of a set of 60 particles that initially
lie on the negative $x$ axis within the horseshoe orbit region exterior to the orbit of the planet ($r > \rp$). The torque (the derivative of the curve)
shows both types of cancellations: the time rate of change of angular momentum oscillates
due to the orbital symmetry and the amplitude of the oscillations decreases due to phase mixing.

A space-filling continuous set of small particles will be considered to represent coorbital gas.
The disk in this case is dissipationless and pressureless, and the planet is on a fixed circular
orbit. As is the case for a set of particles described above,
 the torque that the gas exerts on the planet approaches zero
on timescales longer than the time for gas to make
successive encounters with the planet. 

 We now consider the torque caused by gas whose density varies with radius
with some imposed initial surface density distribution $\Sigma(r)$. The particle paths
in  Fig.~\ref{f:co-traj} then become streamlines for the gas flow. 
Gas loses angular momentum in going from position $o+$  to $i+$. 
 This angular momentum is continuously gained by the planet.
Similarly, the planet continuously loses angular momentum from
the gas that passes from $i_-$ to $o_-$.
The gas spends a relatively short time 
in transition between $r_{\rm{i}}$ and $r_{\rm{o}}$ compared
to the time it spends between encounters  given by equation (\ref{tau}),
where $x = \rp - r_{\rm{i}}$.
We consider here the torque exerted on the planet on timescales
longer than the transition timescale, but shorter than the  timescale between encounters
with the planet. 

As gas in the coorbital region approaches from inside the orbit of the planet, it 
carries a mass flux (defined to be positive) that we estimate as
\begin{equation}
\dot{M}_{\rm i} \sim \Sigma_{\rm i} |\Omega_{\rm i} - \Omega_{\rm p}| r_{\rm i} w_{\rm i},
\label{Mdoti}
\end{equation}
where $w_{\rm i}$ denotes the radial extent (width) of the coorbital horseshoe region
interior to the planet's orbital radius. Quantities $\Sigma_{\rm i}$, $\Omega_{\rm i}$,
and $r_{\rm i}$ denote the values of the surface density, angular velocity, and orbital radius, respectively,
at position $i-$. This position lies
at an intermediate radius within
the horseshoe orbit region and interior to the planet's
orbit. Gas at this position is executing circular motion, just ahead of an encounter with the planet.
Similarly, for coorbital gas outside the orbit of the planet, the mass flux
 is estimated by
\begin{equation}
\dot{M}_{\rm o} \sim \Sigma_{\rm o} | \Omega_{\rm o} - \Omega_{\rm p}| r_{\rm o} w_{\rm o},
\label{Mdoto}
\end{equation}
where the subscripts denote the values at
a point $o+$ that lies at an intermediate radius within the horseshoe orbit region outside
the orbit of the planet. 

Due to the orbital geometry, we have that region widths satisfy
$w_{\rm i} \sim w_{\rm o} \sim r_{\rm o} - \rp \sim \rp-r_{\rm i}$. 
This gas interacts with the planet and undergoes a change in angular momentum  
per unit mass  as it flows from position
$i-$ to $o-$
\begin{equation}
\Delta J = r_{\rm o}^2 \Omega_{\rm o}^2 -   r_{\rm i}^2 \Omega_{\rm i}^2 \sim \rp \Omega_{\rm p}^2 w > 0,
\end{equation}
where $w =  w_{\rm o} + w_{\rm i}  \sim r_{\rm o} - r_{\rm i}$. 
In going from $o+$ to $i+$, the gas undergoes an equal and opposite change in angular momentum, $- \Delta J$.
The flows from outside and inside the
orbit of the planet impart a torque on the planet. This torque is equal and opposite to the rate of change
of angular momentum of gas resulting from the two contributing mass fluxes
\begin{eqnarray}
T_{co} &=& (\dot{M}_{\rm o} - \dot{M}_{\rm i} ) \Delta J \\
           &\sim&  \dot{M}_{\rm o} \left(1 - \frac{\dot{M}_{\rm i}}{\dot{M}_{\rm o}} \right) \rp \Omega_{\rm p}^2 w.
\label{Tco1}
\end{eqnarray}

We need to be careful in evaluating the ratio of the mass fluxes in the above equation,
since departures from unity are critical.
From equations (\ref{Mdoti}) and (\ref{Mdoto}),  the ratio of the mass fluxes is given by 
\begin{equation}
\frac{\dot{M}_{\rm i}}{\dot{M}_{\rm o}} =  \frac{\Sigma_{\rm i}}{\Sigma_{\rm o}} \left( \frac{|\Omega_{\rm i} - \Omega_{\rm p}| r_{\rm i} w_{\rm i}}{|\Omega_{\rm o} - \Omega_{\rm p}| r_{\rm o} w_{\rm o}} \right).
\end{equation}
{\em Ward, 1991} showed that the term in parenthesis on the right-hand side is equal to 
$B_{\rm o}/B_{\rm i}$, 
where $B(r)$ is the Oort constant defined through $2 r B = d (r^2 \Omega)/dr$ and 
$B(r)=\Omega(r)/4$ for a Keplerian disk. 
We then obtain that
\begin{equation}
\frac{\dot{M}_{\rm i}}{\dot{M}_{\rm o}} =  \frac{\Sigma_{\rm i}}{\Sigma_{\rm o}} \frac{B_{\rm o}}{B_{\rm i}} .
\label{mdotr}
\end{equation}
Using equation (\ref{Mdoto}), we then
approximate $\dot{M}_{\rm o}$ appearing in the first term on the right-hand side of equation (\ref{Tco1})
as
\begin{equation}
\dot{M}_{\rm o} \sim  \Sigma \Omega_{\rm p} \rp w^2.
\label{Mdota}
\end{equation}
Applying equations (\ref{Tco1}), (\ref{mdotr}), and (\ref{Mdota}),
we then obtain an expression for the coorbital torque  as
\begin{eqnarray}
T_{\rm{co}} 
     & \sim & \Sigma \Omega_{\rm p}^2 \rp w^3  \left ( \frac{\Delta \Sigma}{\Sigma}-   \frac{\Delta B}{B} \right), \\
    & \sim & \Sigma \Omega_{\rm p}^2 w^4 \frac{d \log{(\Sigma/B)}}{d \log{r}},
  \label{Tco}
\end{eqnarray}
where $\Delta \Sigma = \Sigma_{\rm{o}} - \Sigma_{\rm{i}}$ and 
 $\Delta B = B_{\rm{o}} - B_{\rm{i}}.$
 
Quantity $B/\Sigma$ is sometimes called the vortensity, since
$B$ is  half the vorticity, that is half the curl of the unperturbed velocity $r \Omega \bf{e_{\theta}}$.
The coorbital torque then depends on the gradient of the vortensity.
Recall that in this situation, the torque is not permanent. It oscillates and decays
to zero, due to phase mixing, as discussed above. 
The coorbital torque drops to zero or is said to be saturated on long timescales. The gas density in the coorbital region becomes modified over time so that
the vortensity becomes constant and the torque approaches zero.

Gas that flows on the closed streamlines shown in Fig.~\ref{f:co-traj} 
can impart a change in angular momentum on the planet that is at most about equal to
the gas angular momentum of the region times its fractional width, $w/\rp$ 
(that is, $\Sigma \Omega_{\rm p}  \rp^2 w^2$). This
 angular momentum change is due to the
rearrangement of gas within the horseshoe orbit region from some initial state.
In this model, gas is trapped within
the coorbital region and cannot exchange angular momentum with the large reservoir
in the remainder of the disk. 
Maintaining a coorbital torque over long timescales requires the action of a process 
that breaks the symmetry of the streamlines and permits exchange with the remainder of the disk. One such process involves the disk
turbulent viscosity. The disk viscosity acts to establish a 
characteristic density distribution, as discussed in the Introduction. 
Sufficiently strong turbulence can overcome the effects
of phase mixing that saturate the torque. The torque is then given by equation (\ref{Tco}),
where the density distribution $\Sigma(r)$ is determined by turbulent 
viscosity in the viscous accretion disk. The vortensity gradient is then maintained at a
nonzero value.
The angular momentum changes within the coorbital region are transfered to the remainder of the disk
through torques associated with frictional stresses caused by disk turbulent
viscosity (viscous torques).  In this way, a steady state torque can be exerted on the planet
over timescales much longer than the libration timescale $\sim \rp P/w$.

We compare the coorbital torque with the net torque
due to particles  outside the coorbital region in a pressureless disk, as we discussed in Section 2.2.
Taking $w \sim R_{\rm H}$ in equation (\ref{Tco}), we see that the unsaturated
 coorbital
torque (taking $ |d \log{(B/\Sigma)}/d \log{r}| \sim 1$ and nonzero) is comparable to the 
 torque due to particles outside the coorbital region in equation (\ref{Tnp}). It can be shown
that this is also true for a disk where pressure effects are important,
where $H > R_{\rm H}$. For
such a disk, the coorbital
torque  is generally of the same order
as the net torque from the region outside the coorbital zone, equation (\ref{Tp0}).
The direction of the coorbital torque contribution
depends on the sign of the vortensity gradient.
For the minimum mass solar nebula model,
$\Sigma \propto r^{-3/2}$ and $B \propto r^{-3/2}$.
Consequently, the  coorbital torque
is zero in that case. With more general density distributions of the form
\begin{equation}
\Sigma \propto r^{-\beta}
\label{beta}
\end{equation}
and for smaller values of $\beta < 3/2$,
the coorbital torque on the planet is positive.

We have discussed how the disk turbulent viscosity can cause a torque to
be exerted in the coorbital region over long timescales. But if the amount of viscosity is too
small, the phase mixing process discussed above can dominate and cause the torque to saturate.
For the turbulent viscosity to prevent torque saturation,
the timescale for turbulence to affect the density in the coorbital region needs to be shorter
than the libration timescale on which the torque
would drop in a nonviscous disk. This criterion imposes a constraint
on the level of viscosity required. So although they are potentially powerful,
 coorbital torques are not guaranteed to play
a role in planet migration in all cases. 
 
 Although the coorbital torque is of the same order as the net torque
 due to material outside the orbital zone, more detailed calculations as described
 in the next section show it  typically does not dominate the total torque, 
 at least for simple disk conditions. 
 
 Although the description here is self-consistent,
coorbital torques are less well understood than the torques involving noncoorbital material
we considered in Section 2.2. 
They are more complicated because they are subject to saturation.
The asymmetry
in the coorbital region that prevents torque saturation can in principle be produced by effects other than turbulence, such as the
migration of the planet itself (see Section 3.2). Furthermore, recent simulations suggest that 
the thermodynamic
state of the gas, not considered in this model,
  can modify the coorbital torque and possibly result in outward migration, as is discussed in Section 3.2.
 
\bigskip
\noindent
\textbf{ 2.4 Type~I Migration}
\bigskip

The torques arising from gas that lies somewhat inside 
the orbit  of the planet $r < \rp$ (but outside the coorbital region) act to cause outward migration, while
torques arising from gas outside the orbit of the planet act to cause inward migration, as discussed in Section 2.2. 
The net migration torque was estimated in equation (\ref{Tp0}). But due to
the inaccuracy of the approximations,  the sign
of the torque and its dependence on gas properties,
such as the density and temperature distributions, could not be determined. More accurate analytic calculations
of the torque involve solving the fluid equations for gas subject to the gravitational perturbing
effects of the planet. Both the weakly perturbed region (Section 2.2) and coorbital region
(Section 2.3) are analyzed.  The calculations described in this section
assume a simple disk model. By a simple model we mean 
that the disk has smooth disk density distribution, in the absence of the planet,
and a smooth and fixed temperature distribution, even in the presence of the planet.

Consider a cylindrical coordinate
system centered on the star $(r, \theta, z)$.
In these calculations, the perturbing potential of a circular orbit planet is expanded in a Fourier series as
\begin{equation}
\Phi(r,\theta, t) = \sum_m \Phi_m(r, \rp)  \cos{[m (\theta - \Omega_{\rm p} t) ]},
\label{Phime}
\end{equation}
where $m$ is a nonnegative integer, and $\Omega_{\rm p}$
is the orbital frequency of the planet. Such an expansion is possible because the potential is
periodic in azimuth and periodic in time with the planet's orbital period. In addition, at fixed
radius $r$ (and fixed $\rp$),
 the potential depends only
on the relative azimuth between that of a point $\theta$ and  that of the planet $\Omega_{\rm p} t$, that is 
$\theta - \Omega_{\rm p} t$.
For each azimuthal wavenumber $m$, the gas response is calculated by means of linear theory.
There is a torque associated with each $m$-value due to the effects of the density
perturbation for each $m$.
The sum of these torques determine the net torque on the planet.

 The results show that
 the gas response is dominated by the effects of resonances ({\em Goldreich \& Tremaine}, 1979, 1980).
 There are two types of resonances that emerge: Lindblad resonances and corotational
 resonances. They occur in the regions described in Sections 2.2 and 2.3, respectively.
 Such resonances also arise in the theory of the spiral structure of galaxies and planetary rings.
 
 The gas response to a particular Fourier potential component $m$ 
 in equation (\ref{Phime}) is strong at particular radii $r_m$ where the Lindblad resonance condition is satisfied.
 The Lindblad resonances occur for gas that periodically passes by the planet. 
 They correspond to the so-called mean motion resonances of particles in celestial mechanics.
 For a given $m$-value, the forcing frequency experienced by a particle is  the time
 derivative of the argument of the $\cos$ function in equation (\ref{Phime}) along a particle's
 path $\theta(t)$.
 The forcing frequency is then given by $m(\Omega(r) - \Omega_{\rm p})$.
 Consider a particle in a circular orbit about a central mass.
 If the  particle is momentarily slightly perturbed, it oscillates radially at a frequency called the epicyclic
 frequency, often denoted by $\kappa(r)$. This frequency is the free oscillation
 frequency of the particle, analogous to the free oscillation frequency of a spring
 in a simple harmonic oscillator.
 For a Keplerian disk, the radial frequency $\kappa(r)$ is equal
 to the circular frequency $\Omega(r)$. (This is why noncircular Keplerian orbits
 are closed ellipses in the inertial frame.) 
 Lindblad resonances occur wherever the absolute value of the forcing frequency matches the free oscillation frequency.
 For a Keplerian disk, the Lindblad resonance condition,  $|m(\Omega - \Omega_{\rm p})| = \kappa$,
simplifies to
 \begin{equation}
\Omega(r_m) = \frac{m \Omega_{\rm p}}{m \mp 1},
\label{LRK}
\end{equation}
where $\Omega(r)$ is given by equation (\ref{omk}).

\begin{figure}
 \epsscale{1.0}
 \plotone{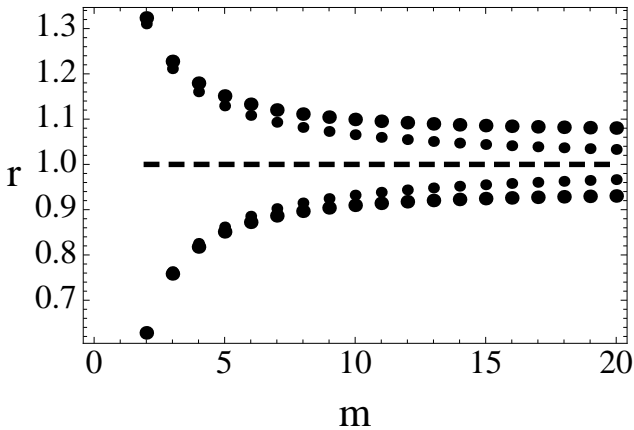}
  \caption{\small Radius of a Lindblad resonance in units of $\rp$ as a function
  of azimuthal wave number $m$. Points below (above) the dashed  line are for inner (outer) Lindblad resonances. The dashed line is the location of the planet.
The two sets of smaller dots are for a 
  Keplerian disk with radii given by equation (\ref{LRK}). The resonances get closer to the planet
  with increasing $m$.
  The two sets of larger
   dots 
  account for azimuthal pressure effects in a disk with thickness to radius ratio $H/r = 0.1$.   
   With these pressure effects included, the resonances maintain a fixed
  separation from the planet at high $m$.
 }
 \label{f:res-r}
 \end{figure}

\begin{figure}
 \epsscale{1.0}
 \plotone{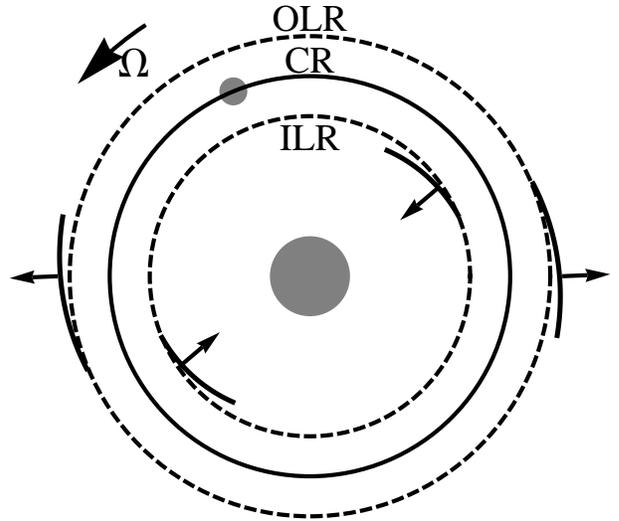}
  \caption{\small Schematic of acoustic (pressure) wave propagation in a gas disk involving
  an inner Lindblad resonance (ILR), outer Lindblad resonance (OLR), and corotation  resonance (CR).
  Spiral waves launched at Lindblad resonances propagate 
   away from the orbit of the planet.
  The region near the CR, between ILR and OLR, is evanescent (nonpropagating). 
}
\label{f:wave-prop}
 \end{figure}

For each $m$-value there are two Lindblad resonances. An outer Lindblad resonance (OLR) occurs
outside the orbital radius of the planet, for which the plus sign is taken in 
the denominator of equation (\ref{LRK})
and an inner Lindblad resonance (ILR), for which the minus sign is taken (see Fig.~\ref{f:res-r}).
At each Lindblad resonance, spiral waves are launched that propagate
away from the resonance (see Fig.~\ref{f:wave-prop}). 
These waves are similar to acoustic waves, but are modified
by the disk rotation. One can associate an angular momentum and energy with the waves.
 They carry energy and angular momentum
from the planet, resulting in a torque on the planet. Various processes cause these
waves to damp as they propagate. As the waves damp, their angular
momentum is transfered to the disk.  This situation
is reminiscent of the case of ocean waves. They are generated
by wind far from land, but undergo final decay when
 they break at the shore and can cause irreversible 
 changes.  Similarly, the waves in disks can modify the 
underlying disk density
distribution $\Sigma(r)$ and open gaps as they damp, although the disk
turbulence can wash out these effects.

At the corotation resonance in a Keplerian disk, the condition is simply that 
\begin{equation}
\Omega(r_m) = \Omega_{\rm p},
\label{CRK}
\end{equation}
or $r_m=\rp$ for all $m$.
That is, the corotation resonance occurs at the orbital radius of the planet.
It lies in the region of orbits shown in Fig.~\ref{f:co-traj}.
The corotation resonance also creates disturbances in the disk, but the disturbances do
not propagate. Instead they are evanescent and remain trapped within a radial
region whose size is of order the disk thickness $H$. 

The analytic calculations determine a torque
for each $m$-value at the inner and outer Lindblad resonances and the corotation resonance.
The strength of the Lindblad torques increases for resonances that lie closer to the
planet. The closer resonances occur for higher $m$-values.
But due to effects of pressure, the resonance condition (\ref{LRK}) breaks down close to the
planet 
and the torque reaches a maximum value for $m_{\rm cr} \sim r/H$ where the resonance
is at a distance of order the disk thickness from the planet  (see Fig.~\ref{f:res-r}). 
Three dimensional
effects also weaken the torque close to the planet. The reduction arises because the gas
is spread over the disk thickness $H$ (as was discussed in Section 3.2).
The critical $m$-value of 
$m_{\rm cr}$ is sometimes called
the torque cutoff. This reduction of torque for $m > m_{\rm cr}$ implies that the sum of the torques
over all $m$-values remains finite. In this way, these calculations avoid the
 the singularity  that we encountered in equation (\ref{Tnet}) involving $\Delta r$.

The results of these calculations show that planets have a definite
tendency to migrate inward ({\em Ward}, 1997). For a given $m$-value, the torque is stronger
at the OLR than the ILR in a  disk, even if we ignore the asymmetries caused by density
and temperature variations. 
There are several contributions to this density- and temperature-independent effect.
One such contribution can be seen
by noticing that for a given $m$, the OLRs lie slightly closer to the planet's orbital radius than ILRs in equation
(\ref{LRK}).
This asymmetry, as well as other density- and temperature-independent effects, enhances
the inward migration effects caused by OLRs  over the outward migration effects
that are caused by ILRs. 
With simple disk structures 
(smooth disk density distributions and 
smooth, fixed temperature distributions), such effects often dominate the overall
torque on the planet for typical density and temperature distributions.

Density and temperature effects can also play an important role
in determining the Lindblad torque. 
The surface density distribution in radius influences the torque balance in at least two ways.
It affects the amount of gas at each resonance and it affects the radial pressure force.
These two effects typically act in opposition to each other and partially cancel. A declining surface density distribution
with radius, as is typical, places more gas at the
ILRs than the OLRs. This effect provides an outward torque contribution, since the ILR torques are enhanced. 
Pressure effects
typically cause the gas at orbital radius close to $\rp$ to  rotate a little more slowly than the planet, as discussed in Section 2.1.
Although the primary effect here is gravitational and not gas drag as in Section 2.1, the gravitational torque
on the planet again
produces a qualitatively similar effect as drag and acts to cause inward migration.
This effect can also be understood in terms of a slightly decreased angular velocity $\Omega(r)$ 
(below the Keplerian rate) in
equation (\ref{LRK})
that causes both the ILRs and OLRs to move inward. The ILRs are then shifted
away from the planet, while the OLRs  are shifted towards it. 
The effect of pressure is to favor the OLRs, which contribute to inward migration.

The temperature distribution in radius also has multiple effects on the Lindblad torque balance.
It affects the gas radial pressure force in a similar way that the density
distribution does, as just described. The pressure effects of a radially declining 
temperature then contribute to inward migration.
Temperature variations also modify the torque cutoff $m_{\rm cr} \sim \rp/H$, since 
disk thickness $H$ depends on the gas sound speed and therefore temperature (see equation (\ref{H})).
The OLRs, being at a lower temperature (smaller $H$) than the ILRs, have a higher cuttoff $m_{\rm cr}$.
This effect favors
the OLRs and its inward torques.
Both effects of a declining temperature with radius, also generally expected,  
contributes to inward migration.

Detailed 3D linear analytic calculations of the Type I migration
rates have been carried out by {\em Tanaka et al}, 2002.
They assumed that the gas sound speed is strictly constant in radius.
That is, the gas temperature is assumed to be unaffected by the planetary
perturbations.
For the case of saturated (zero) coorbital corotation
torques, where only differential Lindblad torques are involved,
the torque on the planet is given by
\begin{equation}
T =  -\left(2.34 - 0.10\,\beta \right)
    \Sigma(\rp) \, \Omega(\rp)^2 \, \rp^4
    \left(\frac{M_{\rm p}}{M_{\rm s}} \right)^2 \left(\frac{\rp}{H}\right)^2,
\label{tan-sat}
\end{equation}
where $\beta$ is given by equation (\ref{beta}).\footnote{A higher sensitivity to density gradients (more negative coefficient of $\beta$) 
was found in the analysis by {\em Menou \& Goodman}, 2004.}
The torque on the planet
resulting from the action of both Lindblad and (unsaturated) 
coorbital corotation resonances is given by
\begin{equation}
T =  -\left(1.36+0.54\,\beta \right)
     \Sigma(\rp) \, \Omega(\rp)^2 \, \rp^4
    \left(\frac{M_{\rm p}}{M_{\rm s}} \right)^2 \left(\frac{\rp}{H}\right)^2.
 \label{tan}
\end{equation}

These migration rates are consistent with
the estimate in equation (\ref{Tp0}).  Numerical values for migration timescales
based on equation (\ref{tan}) are plotted in Fig.~\ref{f:mig-time}.
As mentioned above, temperature 
variations can also contribute to planet migration, but are not included in this treatment.
Other results (such as the 2D calculation by {\em Ward,} 1997) suggest that the temperature variation 
should provide an additional term within the first parenthesis on the right-hand sides of
 equations (\ref{tan-sat}) and (\ref{tan}) that is   $\sim \chi$, where 
 \begin{equation}
 T(r) \propto r^{-\chi}
 \label{chi}
 \end{equation}
 is the temperature distribution with radius  (not to be confused with torque).
 We typically expect in a disk that $0< \beta < 1.5$ and that $ 0.5 \la \chi < 0.75$.
 Consequently,  the migration is predicted to be inward and of order the rate
 that would be obtained even if density and temperature gradients ($\beta$ and $\chi$)
 were ignored.  But, such gradients may take
 on different values than assumed here and potentially play an important
 role in slowing migration, as will be discussed in Section 3.1. In addition, some further effects,
 such as gas entropy gradients,
can modify the coorbital torque. 
Recent studies have suggested that such thermal effects could change Type~I migration and may
even lead to outward migration (Section 3.1; {\em Paardekooper \& Mellema}, 2006).      
 
 In comparing the saturated coorbital torque (equation (\ref{tan-sat}))
  to the unsaturated coorbital torque (equation (\ref{tan})), we see that 
  the effects of the coorbital torque reduce the inward migration rate for $\beta < 1.5$ and 
 nearly vanish for
 $\beta = 1.5$. This behavior is consistent with equation (\ref{Tco}), using the
 fact that $B \propto r^{-1.5}$. 
The coorbital
 torque is then positive for $\beta < 1.5$  and is  zero 
 for $\beta=1.5$.
  
\begin{figure}
 \epsscale{2.0}
 \plottwo{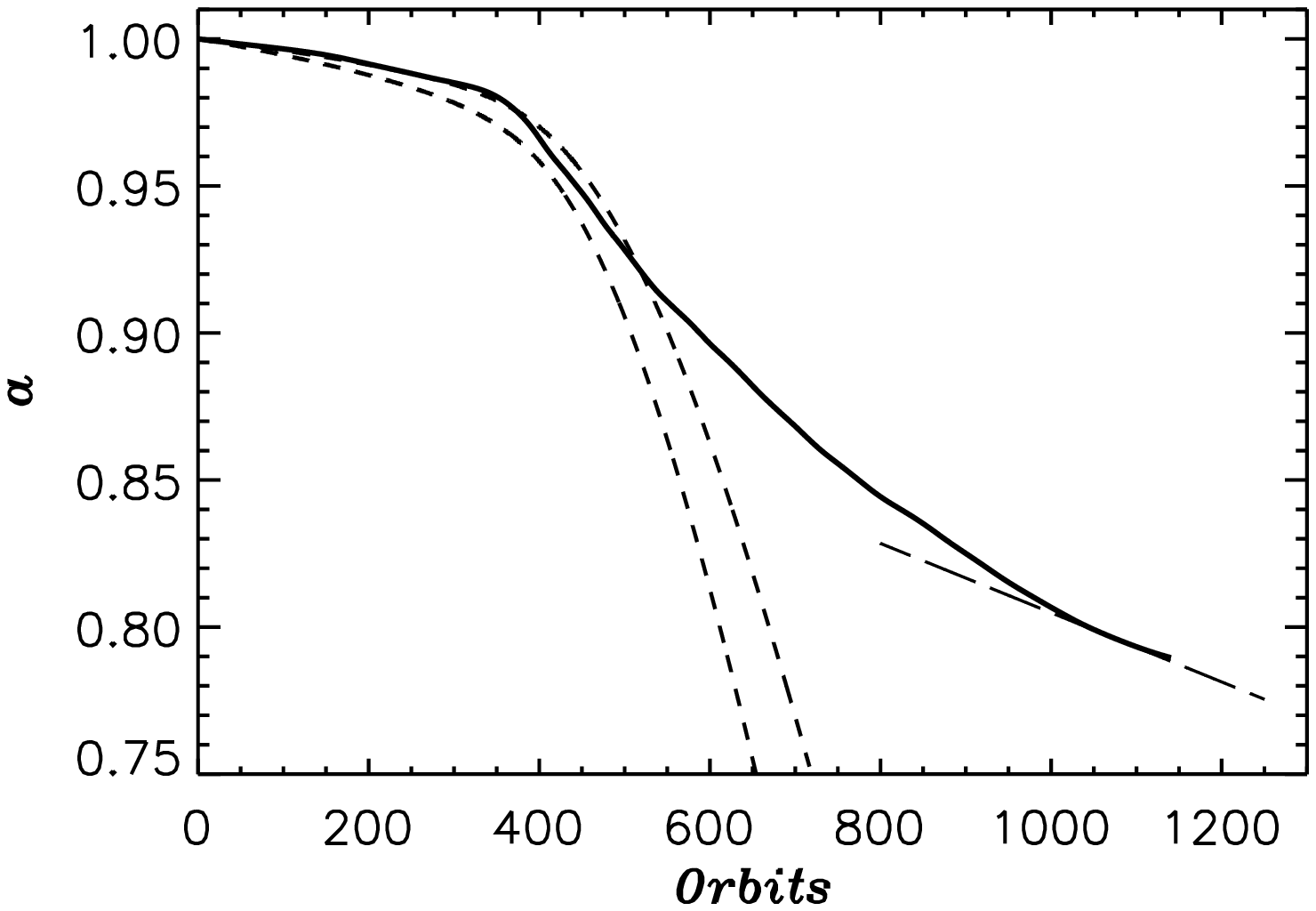}{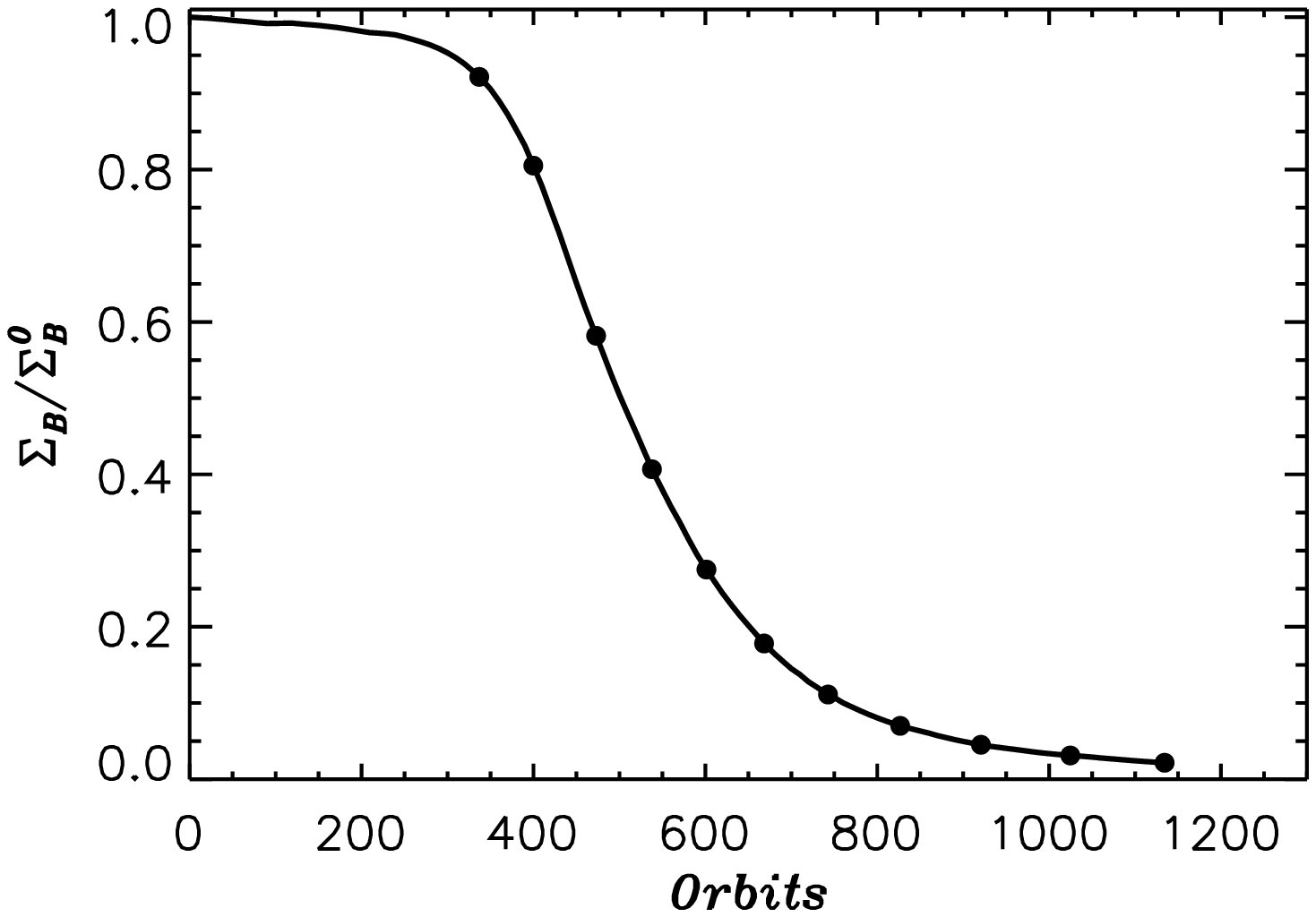}
  \caption{\small Migration
  of a planet orbiting a solar mass star and undergoing growth via gas accretion.  The disk parameters are similar to those in Fig~\ref{f:mig-time}.
 Top: The vertical axis is the orbital radius in units of the initial
  orbital radius $5.2 \AU$. The horizontal axis is time in units of the
  initial orbital period $12 \y$.
 The solid curve is the result of 3D hydrodynamical simulations.
The lower and upper short dashed curves are based on equations (\ref{tan-sat})  
and  (\ref{tan}) respectively, applied to a planet of variable mass. 
   The long dashed curve corresponds to migration on the disk viscous timescale.
   Bottom:  Average disk density near the planet relative to the 
initial value as a function of time. The density is averaged over a band of radial width 
$2H$ centered on the orbit of the planet and is normalized by its initial value. The first solid circle marks the time when 
 $M_{\rm p} = 16.7 M_{\oplus}$, subsequent circles  occur at  integer multiples of  
$33 M_{\oplus}$.
The planet initially follows the predictions of Type~I migration theory in the top panel while there is no substantial
   gap in the disk (no drop in the curve on the bottom figure). After gap opening, the planet follows Type~II migration. 
   Obtained from {\em D'Angelo \& Lubow,} 2008. }
   \label{f:mig-acc}
 \end{figure}

Nonlinear 3D hydrodynamical calculations
have been carried out to test the migration rates, under similar
disk conditions (in particular, local
isothermality) used to derive the analytic model.
Figs.~\ref{f:mig-time} and \ref{f:mig-acc} show that the migration
rates agree well with the expectations of the theory.

\begin{figure}
 \epsscale{1.0}
 \plotone{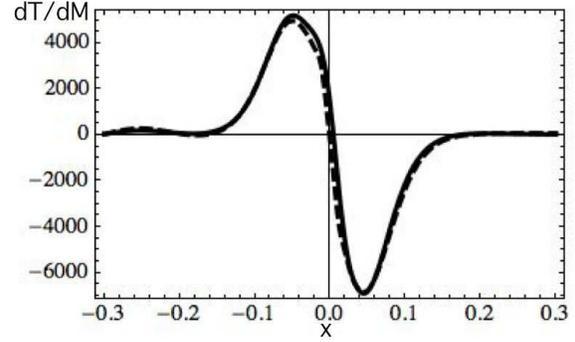}
  \caption{\small Scaled torque per unit disk mass on the planet as a function of radial
         distance from the planet's orbital radius 
          based on 3D simulations. The horizontal scale is in units of $\rp$ and the
         vertical scale is in units of $G M_{\rm s} (M_{\rm p}/M_{\rm s})^2/\rp$. 
         The solid and long-dashed curves
         are for 
         $1\, M_{\oplus}$ (1 Earth mass) and $10\,M_{\oplus}$ 
         mass planets, respectively, that orbit a solar mass star.
         The disk parameters are $H/r=0.05$ and $\alpha= 0.004$
         for both the cases. According to linear theory, these two curves should
         overlap.
         Torque distributions are averaged over one orbital period.
        Figure based on {\em D'Angelo \& Lubow}, 2008.
 }
 \label{f:dTdM}
 \end{figure}

We examine the comparison between simulations
and theory in more detail by comparing torque
distributions in the disk as a function of disk radius.
We define  the distribution of torque on the planet per unit
disk mass as a function of radius as
$dT/dM(r) = 
1/(2 \pi r \Sigma(r))\, dT/dr(r)$. 
Fig.~\ref{f:dTdM} plots  $dT/dM$ (scaled as indicated in the figure) as a function
of radial distance from the planet based on 3D simulations. 
The distributions show that the region interior (exterior) to the planet provides
a positive (negative) torque on the planet, as predicted in equations (\ref{To})
and (\ref{Ti}) for a particle disk. Also, the integrated total torque is negative, implying inward migration. 
The theory predicts
that the torque density peak and trough 
occur at distance from the
planet $r- \rp=\Delta r \sim \mp H$, where the torque cutoff takes effect. 
For the case plotted in Fig.~\ref{f:dTdM} that adopts $H = 0.05 r$, the predicted 
locations  agree well with the locations
of the peaks and troughs in the figure. 
Furthermore, the torque and torque density should
scale with the square of the planet mass.
Although the vortensity gradient is not small, since $\Sigma/B \sim r$,
we do not find large contributions to the torque density from the coorbital region,
as is also expected.
As seen in Fig.~\ref{f:dTdM}, this expectation
is well met for the two cases plotted.

 \bigskip
\noindent
\textbf{ 2.5 Type~II migration}
\bigskip

The planet's tidal torques 
cause material interior to the orbit of the planet (not in the coorbital region) to
lose angular momentum and material exterior to the orbit of the planet
to gain angular momentum, as we saw in Section 2.2. The torques then act to clear a gap about
the orbit of the planet. 
As discussed in the Introduction,
Type~II migration occurs when the planet mass
is sufficiently large that tidal forces cause a gap
to clear about the orbit of the planet  ({\em Lin \& Papaloizou}, 1986). 
The tidal torques on the 
disk interior or exterior to the planet
are estimated by equation (\ref{To}) with $\Delta r \sim H$. 
The tidal torque on the disk that acts to open the gap is then estimated as
\begin{equation}
T_{\rm o} \sim \Sigma \, \Omega_{\rm p}^2 \, \rp^4 \left( \frac{M_{\rm p}}{M_{\rm s}} \right)^2 
        \left( \frac{\rp}{H} \right)^3.
\label{Tt}
\end{equation}

Turbulent viscosity acts to close the gap. 
The effects of turbulence are often described in terms of a kinematic viscosity $\nu$
that is parameterized in the so-called $\alpha$-disk model  as
\begin{equation}
\nu = \alpha c H,
\label{alpha}
\end{equation}
where $\alpha$ is a dimensionless number, $c$ is the gas sound speed, and $H$ is the
disk thickness, described by equation (\ref{H}) ({\em Shakura \& Sunyaev,} 1973). 
The value of the key quantity $\alpha$ is expected to be less than unity because we presume
that the turbulent motions are slower than sonic and that the characteristic length scales for the turbulence
are smaller than the disk thickness.

The turbulent viscosity provides a torque to close the gap
\begin{equation}
T_{c} \sim \frac{M_g \Delta J_g }{t_g},
\label{Tv1}
\end{equation}
where $M_g$ is the mass supplied by the disk in closing the gap, $\Delta J_g$ is the 
change in the disk angular momentum per unit mass required to close the gap, and $t_g$ is the
time for gap closing.
We estimate these quantities as
\begin{eqnarray}
M_g &\sim& \Sigma \rp w, \\
\Delta J_g &\sim& \rp \Omega_{\rm p} w,\\
t_g &\sim& \frac{w^2}{\nu},
\end{eqnarray}
 where $w$ is the radial extent of the gap, $\Sigma$ is the disk density just outside the gap,
 and $\nu$ is the disk kinematic turbulent viscosity.
The gap closing timescale estimate $t_g$ is based on the idea that the viscosity acts on a radial diffusion timescale
across a distance $w$, which then implies the quadratic dependence on $w$.
Applying this last set of relations to equation (\ref{Tv1}), we obtain that
\begin{equation}
T_{\rm c} \sim \Sigma \, \rp^2 \, \Omega_{\rm p} \, \nu,
\label{Tv}
\end{equation}
independent of $w$.

\begin{figure}
 \epsscale{1.0}
 \plotone{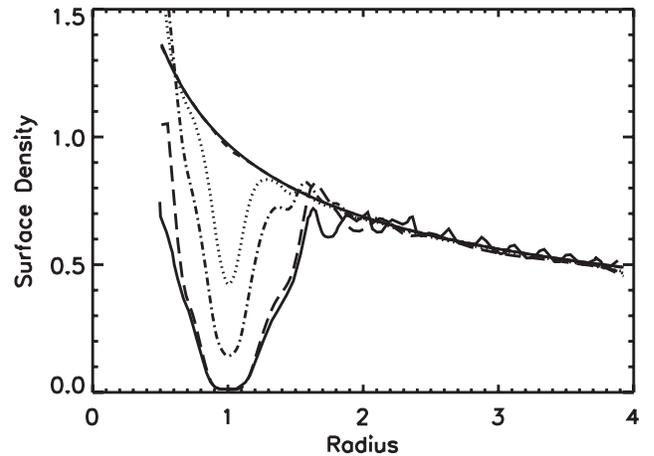}
  \caption{\small Azimuthally averaged disk surface density normalized
  by the unperturbed value at radius $\rp$ as a function of $r$ in units of $\rp
  =5.2 AU$. The central star has a solar mass.
         The disk is simulated in 3D with
          parameters $H/r=0.05$ and $\alpha= 0.004$.
          The density profiles are for planets with masses of 1 (long-dashed),
          0.3 (dot-dashed), 0.1 (dotted), 0.03 (short-dashed), and 0.01 (thin solid) $M_{\rm J}$. Only planets with masses $M_{\rm p} \ga 0.1 M_{\rm J}$ produce significant
          perturbations. The thick solid line is based on a 2D simulation of a $1 M_{\rm J}$ planet
          by Lubow et al (1999). Obtained from {\it Bate et al}, 2003.
           }
 \label{f:densq}
 \end{figure}

In order to open a gap in the disk, the gap opening torque must 
be greater than the gap closing torque, $T_{\rm o} \ga T_{\rm c}$. 
We then obtain a condition on the planet mass required to open a gap
\begin{equation}
\frac{M_{\rm p}}{M_{\rm s}} \ga C_g \left( \frac{ \nu}{\rp^2 \Omega_{\rm p}} \right)^{1/2} 
\left( \frac{H}{\rp} \right)^{3/2},
\label{gap}
\end{equation}
where $C_g$ is a dimensionless number of order unity.
{\em Lin \& Papaloizou, 1986} estimated that $C_g = 2 \sqrt{10}$.
For disk parameters $\alpha = 0.004$ and $H/r = 0.05$,
the predicted gap opening at the orbit of Jupiter about a solar mass star
occurs for planets having a 
mass $M_{\rm p} \ga 0.2 M_{\rm J}$, where $M_{\rm J}$ is Jupiter's mass. This prediction is in
good agreement with the results of 3D numerical
simulations (see Fig.~\ref{f:densq}).

In addition to
the above viscous condition, an  auxiliary condition
for gap opening has been suggested based on the stability
of a gap. 
This condition is to preclude gaps for which steep density gradients
would cause an instability that prevents gap opening.
This  condition, called the thermal condition, is given by the requirement
that the Hill radius, given by equation (\ref{Hill}), be larger than the disk thickness,
$R_{\rm H} \ga H$ ({\em Lin \& Papaloizou}, 1986).
The critical mass for gap opening by this condition is given by
\begin{equation}
\frac{M_{\rm p}}{M_{\rm s}} \ga 3\left( \frac{H}{\rp} \right)^{3}.
\label{gap2}
\end{equation}
For $H/r=0.05$, this condition requires a larger planet mass for gap opening 
than equation (\ref{gap}) for $\alpha \la 0.01$.
A condition that combines both equations (\ref{gap}) and (\ref{gap2})
has been proposed by {\em Crida et al}, 2006. 
Even if both the thermal and viscous conditions are satisfied and $R_{\rm H} \ga H$,
a substantial gas flow may occur
though the gap and onto the planet ({\em Artymowicz \& Lubow,} 1996; {\em Lubow \& D'Angelo,} 2006).

The migration rate of a planet
embedded in a  gap is quite different from the Type~I case
that we have already considered.
A planet that opens a gap in a massive disk, a disk whose
mass is much greater than the planet's mass, would be
expected to move inward, pushed along with the disk accretion inflow. The planet simply
communicates the viscous torques across the gap by means of tidal
torques that balance them. 
The Type II migration timescale is then of order
the disk viscous timescale 
\begin{equation}
t_{\rm vis} \sim \frac{\rp^2}{\nu}
\sim \frac{\rp^2}{\alpha c H}
\sim \left(\frac{\rp}{H}\right)^2 \frac{1}{\alpha \Omega_{\rm p}},
\label{t_vis}
\end{equation}
which is $\sim 10^5$ years  at Jupiter's orbital radius about the Sun for 
$\alpha = 0.004$, $H = 0.05 \rp$, and $\Omega_{\rm p} = 2 \pi/12 \, \,\yi$.
Therefore, the migration
timescale can be much longer than the Type I migration timescale for
higher mass planets that open gaps,
as is found in simulations ($M_{\rm p} > 0.1 M_{\rm J}$ in Figs.~\ref{f:mig-time}
and \ref{f:mig-acc}).
However, this timescale is still shorter than the observationally inferred
global disk depletion timescales $\sim 10^6-10^7$ years.
The actual migration rate may need to be somewhat smaller,
in order to explain the abundant population 
of observed extrasolar 
giant planets beyond 1AU ({\em Ida \& Lin} 2008).


In practice, the conditions for pure Type II migration
are unlikely to be satisfied. The disk
mass may not be very large compared to the 
planet mass and the disk gap may not be fully clear
of material. However, simulations have shown that
to within factors of a few, the migration timescale
matches the viscous timescale, $t_{\rm vis}$. This result holds over a
wide range of parameters, provided that the tidal clearing is substantial
and the disk mass is at least comparable
to the planet mass (e.g.,  Fig.~\ref{f:mig-acc}). 

Another way of understanding the Type II torques
is to recognize that the distribution of tidal torques per unit
disk mass, as seen in Fig.~\ref{f:dTdM}
still approximately applies, even if the disk has a deep gap that
is not completely clear of material. The 
disk density through the gap region adjusts so that 
the planet migration rate is compatible  with the 
evolution of the disk-planet system.

\bigskip
\centerline{\textbf{ 3. OUTSTANDING QUESTIONS}}
\bigskip

We describe some of the major issues involved with 
the theory of planet migration. The topics, descriptions, and references are by no means intended
to be complete. The purpose is to introduce a few of the questions
and some of the suggested solutions.

Planet migration is difficult to calculate for a variety of reasons, as described
near the end of the Introduction. There are technical challenges in determining
the properties of migration torques.
For example, the coorbital torques are subject to saturation by nonlinear feedback effects.
Lindblad torques grow in strength with proximity to the planet. Their contributions
depend on the details of how the torques are limited near the planet. For the simple
disk structures (e.g., moderate turbulent viscosity, fixed and smooth temperature distributions, smooth
density distributions - apart from the density perturbations and gaps produced by the planet), the theory
has been verified by various 
multi-dimensional nonlinear hydrodynamic simulations (e.g., Figs.~\ref{f:mig-time}, \ref{f:mig-acc}, and \ref{f:dTdM}). But such simple disk structures may not arise in real systems. Indeed, the problems with shortness
of planet migration timescales suggests that more complicated disk structures and 
physical processes (such as more complicated thermal effects) may be important. The disk structures influence
migration, since migration torques are sensitive to gradients of disk properties. But we do
not have a complete theory for these disk structures. In addition, migration depends somewhat on the detailed
properties of disk turbulence that are not well understood.

\bigskip
\noindent
\textbf{ 3.1 Limiting/Reversing Type~I Migration}
\bigskip

As seen in Fig.~\ref{f:mig-time}, the 
timescales for Type~I migration, based on the model described in Section 2, are short compared
to gas disk lifetimes of several million years. They become even shorter for a disk
having a mass greater than the minimum mass solar
nebula.  As emphasized earlier, 
the shortness of the timescales for Type~I migration is a serious problem for understanding
how planets form and survive in a gaseous disk.
To be consistent with the ubiquity of extrasolar gas giants
and formation of Jupiter and Saturn, some studies
suggest that the inward Type~I migration rates
must be reduced by more than a factor of 10 ({\em Alibert et al.} 2005; {\em Rice \& Armitage} 2005;
{\em Ida \& Lin} 2008).
One major question is whether there are processes
that could slow the migration. Several
ideas have been proposed. Major uncertainties
about them arise due to our lack of knowledge about the detailed structure
of the disks.  

We briefly discuss below a few of the several suggested mechanisms for
slowing Type I migration.  We also 
discuss some processes that may
reverse migration, resulting in outward migration. 

The fundamental tendency for Type~I to be rapid can be difficult
to cancel without artificially fine tuning some other effect that provides outward torques.
In general, processes that provide
outward torques tend to result in either rapid inward or rapid outward migration. 
However, there are effects that provide an opposing
torque that increases as the planet continues migrating (via  feedback effects
or "traps"). Such  processes can naturally maintain halted migration
once the inward and outward torques are in balance.

{\it Weak turbulence}

\begin{figure}
 \epsscale{1.0}
 \plotone{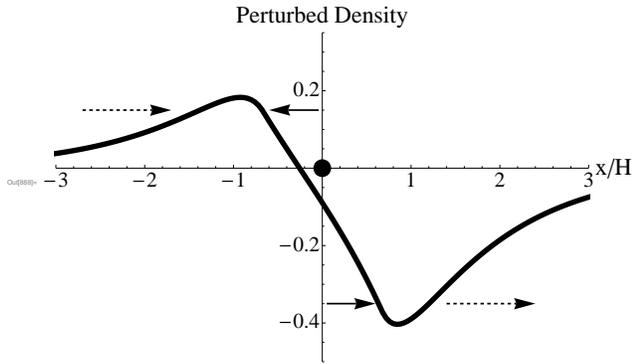}
  \caption{\small 
  Schematic of the axisymmetric density perturbation in arbitrary units
  as a function of radial distance from the planet (shown as dot
  at the origin) in units of disk thickness $H$. The solid lines with arrows
  indicate the velocity perturbation on the gas caused by tidal torques
  on the planet. These velocities are directed away from
  the orbit of the planet. The dashed lines with arrows indicate the gas velocity,
  in the frame of the planet, due to the assumed inward migration of the planet.
  Interior to the orbit of the planet, the velocities due to  tides and migration
  oppose each other and lead to a pile up of material. Outside the orbit
  of the planet, the velocities reinforce each other and lead to a density drop.
           }
 \label{f:feedback}
 \end{figure}

For an inwardly migrating planet,
there is a dynamical feedback effect  associated with the radial motion of the planet that 
raises the axisymmetric gas surface density within a region
interior to the orbit of the planet,
and lowers the density within a region exterior to the orbit ({\em Ward}, 1997), in the same sense as a plow
operating in the radial direction.  The planet pushes material away
from its orbital radius as it migrates. This leads to a pile-up of material
in the direction of its radial motion (see Fig.~\ref{f:feedback}).
The feedback then 
enhances the positive torques that arise in the inner disk and slows inward migration.
For the feedback to exist, the effects
of disk turbulence must be sufficiently weak, otherwise the density
perturbations are erased by the effects of turbulent diffusion.
This  feedback grows with planet mass. Above some
critical planet mass $M_{\rm{cr}}$, the feedback becomes strong enough that the planet can no 
longer migrate.
When migration is halted, the density perturbations are initially mild and there is no substantial gap
(unlike the Type~II migration case).
There is just a sufficient density asymmetry across the orbit of the planet to change the 
competition between inward and outward torques away from favoring the inward torques.
In subsequent evolution, the planet begins forming a gap.

\begin{figure}
 \epsscale{1.0}
 \plotone{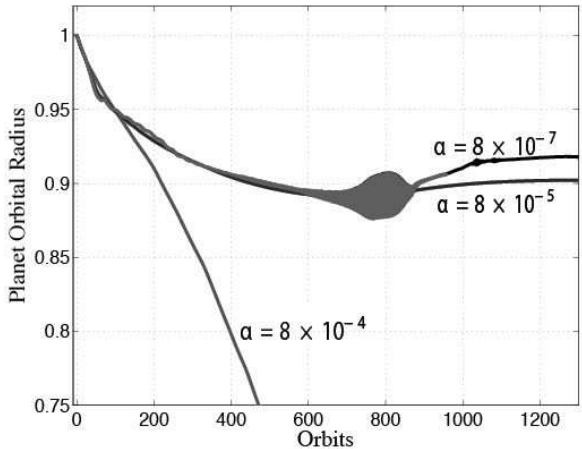}
  \caption{\small Influence of disk viscosity parameter $\alpha$  on the
  migration  of a planet with
  mass $10 M_{\oplus}$ in a disk with $H/r = 0.035$ and  mass
  $\sim 0.1 M_{\odot}$ based on 2D simulations.
  The vertical axis is the orbital radius in units of the initial orbital
  radius. The horizontal axis is the time in units of the initial planet
  orbital period.  For $\alpha = 8 \times 10^{-4}$ the migration
  follows the Type I rate. At the lower values, the migration halts due
  to a feedback effect (see Fig. \ref{f:feedback}). Figure based on  {\em Li et al}, 2008.
           }
 \label{f:lowv}
 \end{figure}

 The value of the critical
planet mass $M_{\rm{cr}}$ depends on several factors such as how far
from the planet the density perturbations occur. 
For disks with $H/r \sim 0.05$ and low turbulent viscosity
$\alpha \la 10^{-4}$, the density perturbations produced by 
shocked density waves in 2D disks cause migration to be sharply reduced at values of 
$M_{\rm{cr}} \sim 10 M_{\oplus}$ (10 Earth masses)
({\em Rafikov}, 2002; {\em Li et al}, 2009; see Fig.~\ref{f:lowv}). 
For disk turbulent viscosity parameter $\alpha \ga 10^{-3}$, Type~I migration proceeds with little reduction,
since the density perturbations are washed out by the effects of disk turbulence.
We have little direct information on what the $\alpha$ value should be.
Some indirect evidence based on observed accretion rates onto young stars
suggests that the average $\alpha \ga 10^{-4}$. Theory suggests that there may be disk regions 
of low $\alpha$ in magnetically unstable disks (the so-called dead zones of low ionization {\em Gammie}, 1996).

{\it Disk Density/Temperature Variations}

The torque on a planet
 depends somewhat on differences in 
properties of the disk across  the planet's orbit, as we have seen in Section 2. 
If the density and temperature 
smoothly decreases in radius at rates typically expected,
the outcome is inward migration (see equation (\ref{tan}) and Fig.~\ref{f:mig-time}).
We expect this 
to be generally true across the disk, but there could be regions where the
density abruptly changes or where the radial temperature gradient is less
negative or is even positive. 
To slow migration, the effects of these gradients would
need to counteract the general trend of inward migration discussed
earlier that is due to other asymmetric effects, such as the asymmetries associated with Keplerian rotation.
It is possible that
rapid changes in disk
properties  with radius could occur as a consequence
of  strong radial variations in disk opacity or turbulent
viscosity.
They can in turn modify the typically adopted gradients of density and temperature that
led to our estimated migration timescales. 
A planet could then experience slowed
migration or even be trapped with
no further inward migration (e.g., {\em Menou \& Goodman}, 2004;
{\em Masset et al}, 2006, {\em Matsumura et al}, 2007).
However, our current knowledge about disk structure is limited.

Disk temperature variations can also be caused by the planet itself.
The planet reduces the disk thickness  $H$
of gas close to it, as a consequence of its gravity. As a result, gas close to and just outside the orbit of the planet
is further exposed to the stellar radiation and experiences additional heating by the star.
Gas just inside the orbit of the planet experiences some shadowing.
This effect acts to decrease the temperature gradient $\chi$ defined in equation (\ref{chi}). Calculations by
 {\em Jang-Condell \& Sasselov}, 2005 have shown that this effect can reduce the Type~I
 migration rates by up to about a factor of 2.

{\it Turbulent Fluctuations}

The effects of disk turbulence are typically described
by means of a turbulent viscosity, e.g., equation (\ref{alpha}).
The numerical simulations  used in Fig.~\ref{f:mig-time} applied
this model. 
Viscosity provides  an approximate description of 
the dynamical effects of turbulent fluctuations
that  are averaged over  the small space and short time scales characteristic of
the turbulence.
But,  the time-dependent, small-scale
density fluctuations can give rise to fluctuating random
torques on the planet that are not described by viscosity.  
Unlike the Type~I torque that
acts continuously in the same direction, the fluctuating
torque undergoes changes in direction on timescales
characteristic of the turbulence
that are short compared with the migration timescale.
The fluctuating torque causes the planet
to undergo something like a random walk. For the effects of the random walk to completely 
dominate over Type~I migration, the 
amplitude of the fluctuating torque must be much larger than
the Type I torque that acts steadily.

The
change in the angular momentum of the planet
due to a random torque $T_{\rm R}$  over some time $t$ is given
by $\sim \sqrt{N} T_{\rm R} t_{\rm R}$, where $N = t/t_{\rm R}$
is the number of fluctuations felt by the planet and
$t_{\rm R}$ is the characteristic timescale for the torque fluctuation,
perhaps of order the orbital period. The planet
experiences many torque fluctuations over the migration timescale, $N \gg 1$.
The change in the planet angular momentum by the Type~I torque $T$
is given by $\sim T t$. For random
torques to dominate, we require $T_{\rm R} \ga \sqrt{N} T$. We take $t$ to be the Type~I timescale of $\sim 10^5 \  \y$
and $t_{\rm R} \sim 10 \, \y$, so that  \ $N \sim 10^4$ and we require $T_{\rm R}$ to be  $ \ga 100 T$.
Torque $T_{\rm R}$ depends linearly on the planet mass,
while the Type~I torque $T$ increases quadratically
with the planet mass and the Type~I migration time $t$ decreases with
the inverse of the planet mass. Therefore, the random torque
is then more important for lower mass planets.
The nature of the random torque depends on the
properties of the disk turbulence, in particular
the amplitude $T_{\rm R}$ and correlation timescale $t_{\rm R}$ for the fluctuating torques, which are currently not
well determined. 

\begin{figure}
 \epsscale{0.75}
 \plotone{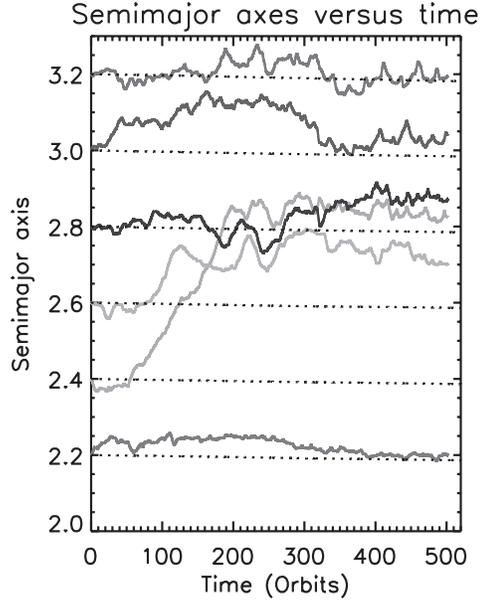}
  \caption{\small 
  Orbital radius  as a function
  of time in orbits for $3 M_\oplus$ planets embedded in a gaseous
  disk based on simulations. 
 The dotted lines plot the migration of planets in a disk without 
 turbulent fluctuations. The solid lines plot the migration 
 of planets in a disk with turbulent fluctuations due to the MHD turbulence.
      The random motions are due to the fluctuating
  torques.
    Obtained from  {\em Nelson,}  2005.
           }
 \label{f:turb-mig}
 \end{figure}

Some global simulations and analytic models
suggest that turbulent fluctuations arising from a magnetic
instability (the magneto-rotational instability {\em Balbus \& Hawley,} 1991) are important for migration of lower mass planets 
({\em Nelson}, 2005; {\em Johnson et al.}, 2006; see Fig.~\ref{f:turb-mig}).
Some other simulations, those done for a small region
of the disk called "box" simulations, suggest that these fluctuating torques  are not effective enough to play an important role in planet migration
({\em Yang et al.}, 2009).  Both the global and box simulations are very computationally demanding
and the current results cannot be regarded as definitive.
The box simulations have higher resolution than the global
ones, since they cover a smaller region. However, they may miss important effects,
if they occur on larger scales
than the box size.
The direct global simulations, such as
those in Fig.~\ref{f:turb-mig},  cover
much less time than the Type~I migration timescale for a low mass mass planet and
consequently may not reveal
the competing effects of unidirectional Type~I torques.

Unlike Type~I migration, the survival of a planet within a disk where fluctuating torques
dominate is described statistically. Over time, there is a smaller probability that planets survive in the disk.
But there is always a nonzero probability of survival.  

However, if turbulent fluctuations are important for migration, then the
eccentricities of planetesimals are pumped up so highly that collisions between
them may result in their destruction rather than accretion 
({\em Ida et al.}, 2008). 
Therefore, although the turbulent fluctuations 
may inhibit the infall of planetary cores into the central
star by migration, they may inhibit the build up of the cores
necessary for giant planet formation in the core accretion model. 

{\it Ordered Magnetic Field}

An ordered magnetic field can affect the nature of
disk-planet interactions. {\em Terquem}, 2003 analyzed
the effects of an ordered (nonrandom) torroidal magnetic field on Type~I planet migration.
The magnetic field introduces additional resonances, called magnetic resonances, that lie closer
to the planet than the Lindblad resonances, whose locations are given by 
equation (\ref{LRK}). As a result, the magnetic resonances can be stronger than the Lindblad ones.
As in the case of the Lindblad resonances, a magnetic resonance located interior (exterior)
to the orbit of the planet causes outward (inward) planet migration. If the magnetic field
strength falls off sufficiently fast with distance from the star, the inner magnetic resonance dominates,
and slowed or even outward migration can occur. For example, in the case that the magnetic
energy density is comparable to the gas thermal energy density in the disk, an outward torque on the planet
can be produced, if the magnetic-to-gas energy density ratio falls off faster than $r^{-2}$. 
How this ratio this varies within a disk is not known,
although such a variation is quite plausible in some situations. If the magnetic field is responsible for the
disk turbulence,
then the torroidal field would contain a fluctuating component that complicates the outcome.

{\it Migration Driven By Nonisothermal Effects in the Coorbital Region}

The disk is heated by the central star and by viscous turbulent dissipation.
Many  studies of disk-planet interactions simplify the disk temperature structure 
to be locally isothermal. In this approximation the temperature distribution depends
on radius in a fixed, prescribed manner. Such a situation can arise
if the optical depth of the disk is low enough that it efficiently radiates
any excess energy due to compression caused by interactions with the planet.
The locally isothermal assumption  is
frequently applied in numerical simulations and was applied in obtaining equation (\ref{tan}). 
The behavior in the isothermal limit tends to suggest that 
coorbital torques do not typically dominate migration (e.g., equation (\ref{tan})).

But, recent work has suggested that nonisothermal 
behavior could have an important effect on coorbital torques and the overall planet migration rate.
The nonisothermal regime has been explored in simulations by {\em Paardekooper \& Mellema}, 2006 
who found that slowed and even outward migration due to coorbital torques may occur in certain regimes. 
Recent studies have suggested that a background radial entropy gradient could play a role
in determining the corotational torque if the gas undergoes adiabatic
changes in its interactions with the planet. This effect may provide an additional contribution
to the coorbital torque beyond the vortensity gradient that
appears in equation (\ref{Tco}).
The slowing/reversing of migration appears to involve the conditions that the disk have a negative radial entropy gradient, 
sufficient viscosity
to avoid coorbital torque saturation, and a thermal timescale that is long enough for the
gas to behave adiabatically as it passes the planet.
The nature of this effect and the conditions required for it to operate
 is an active area
of investigation (e.g., {\em Baruteau \& Masset}, 2008).

\bigskip
\noindent
\textbf{ 3.2 Other Migration Processes Involving a Gaseous Disk}
\bigskip

{\it Runaway Coorbital  Migration, Type~III Migration}

\begin{figure}
 \epsscale{1.0}
 \plotone{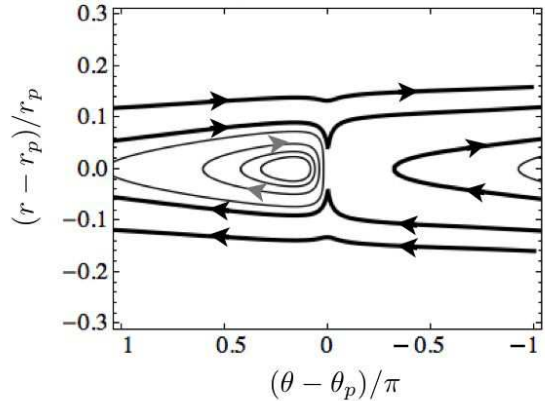}
  \caption{\small Coorbital streamlines near an inwardly
  migrating Saturn mass planet
  $M_{\rm p} = 3 \times 10^{-4} M_{\rm s}$ 
  located at the origin  in the comoving frame of the planet that orbits a solar mass star. 
    The inward migration rate is $0.002 \Omega_{\rm p} \rp$.
Angle $\theta$
  increases to the left. The heavy streamlines are open and pass by the
  planet. The light streamlines are closed and contain trapped material on
  the leading side of the planet
  (streamlines based on {\em Ogilvie \& Lubow}, 2006). 
  The trapped material, acquired at earlier time, is retained from regions further from
  the star. The open streamlines carry
  ambient disk material. 
  In contrast to the
  nonmigrating case in Fig.~\ref{f:co-traj},
  the asymmetry and the density differences between the 
  trapped (retained) and open (ambient)
  gas gives rise to a coorbital torque. }
  \label{f:co-mig}
 \end{figure}

The coorbital torque
will be saturated (reduced to zero) unless some
effect is introduced to break the symmetry of the streamlines, such as turbulent
viscosity (see discussion in Section 2.2). However, planet migration itself 
introduces an asymmetry and could therefore act to prevent
torque saturation.  
(The coorbital torque model presented in Section 2.3 ignored
the effects of planet migration on the horseshoe orbits in
Fig.~\ref{f:co-traj}.) The coorbital torque for a migrating planet could then depend on the rate of migration.
 Under some conditions, the coorbital torque could in turn cause faster migration and in turn a stronger torque,
resulting in an instability and  
a fast mode of migration ({\em
Masset \& Papaloizou}, 2003). 
The resulting migration
is sometimes referred to as Type~III migration. 
To see how this might operate in more detail, we consider 
the evolution of gas trapped in the coorbital region
({\em Artymowicz}, 2004; {\em Ogilvie \& Lubow}, 2006).
For a sufficiently fast migrating planet, the topology
of the streamlines changes with open streamlines flowing
past the planet and closed streamlines containing 
trapped gas (see Fig.~\ref{f:co-mig}).
The  leading side of the planet contains trapped
gas acquired at larger radii, while the gas on the trailing side
is ambient material at the local disk density. The density contrast
between material on the trailing and leading sides of the planet
gives rise to a potentially strong torque. 
The migration timescales 
are of order the Type~I migration timescales. This timescale can be very short
for planets that are massive enough to partially open a gap and would otherwise
not undergo Type~I migration.
The migration timescales
have been found to be as short as 10-20 orbits in the case of a Saturn mass planet
embedded in a cold and massive disk.
 A major question centers around the conditions required
for this form of migration to be effective and therefore whether planets
typically undergo such migration. 

{\it Migration of Eccentric Orbit Planets}

The analysis of migration in Section 2 assumed that planets
reside on circular orbits. This assumption is not unreasonable,
since there are strong damping effects on eccentricity
for a planet that does not open a gap in the disk ({\em Artymowicz}, 1993).
Some eccentricity may be continuously produced by turbulent
fluctuations in the gas, as described above,
or by interactions with other planets.
In general, eccentricity damping is faster than Type~I migration.
For planets that
open a gap, it is possible that they reside on eccentric orbits
in the presence of the gaseous disk. In fact, one model for the observed
orbital eccentricities of extra-solar planets attributes the excitation
of eccentricities to  disk-planet interactions ({\em Goldreich \& Sari}, 2003;
{\em Ogilvie \& Lubow}, 2003). 

A planet on a sufficiently eccentric orbit embedded in a circular  disk
can orbit more slowly at apoastron than the exterior gas 
with which it tidally interacts. Similarly, a planet can orbit more 
rapidly than the tidally interacting gas at periastron. These
angular velocity differences can change the nature
of the "friction" between the planet and the disk discussed in Section 2.2.
For example, at apoastron the more slowly orbiting planet could gain angular momentum
from the more rapidly rotating nearby gas that lies outside its orbit.  
Furthermore, since the planet spends more time at apoastron than periastron,
the effects at apoastron could dominate over effects at periastron.
It is then possible that outward migration could occur for eccentric
orbit planets undergoing Type I migration, assuming
such planets could maintain their eccentricities ({\em Papaloizou}, 2002). 

In  the case of a planet that opens a gap,
simulations suggest that slowed or even outward  migration may occur as the 
planet gains eccentricity from disk-planet interactions ({\em D'Angelo et al},
2006).  The situation is complicated by the fact that
the gaseous disk can gain eccentricity from the planet
by a tidal instability ({\em Lubow,} 1991). For the outward torque to be effective, there
needs to be a sufficient difference in the magnitude and/or orientation between
the planet and disk eccentricities, so that
the planet moves slower than nearby disk gas at apoastron.

{\it Multiplanet Migration}

Thus far, we have only considered single planet systems.
Of the more than 200 planetary systems detected to date by Doppler techniques, over 10\% are found in multi-planet systems 
({\em Butler et al}, 2006). Systems have been found to have orbits
that lie in mutual resonance, typically the 2:1 resonance.
The resonant configurations are likely to be the result of 
convergent migration, migration in which the separation
of the orbital radii decreases in time. This process occurs 
as the outer planet migrates
inward faster than the inner planet.  Planets can become
locked into resonant configurations and migrate
together, maintaining the planetary orbital frequency ratio
of the resonance (see also Chapter 10). The locking can be thought of as a result of
trapping the planets within a well of finite depth.
Just which resonance the planets become
locked into depends on their eccentricities and the relative rate of migration
that would occur if they migrated independently.
As planets that are initially well-separated come closer
together,  they lock into the first resonance that provides
a deep enough potential to trap them against the effects
of their convergence. We discuss below some consequences
of resonant migration.

\begin{figure}
 \epsscale{1.5}
 \plottwo{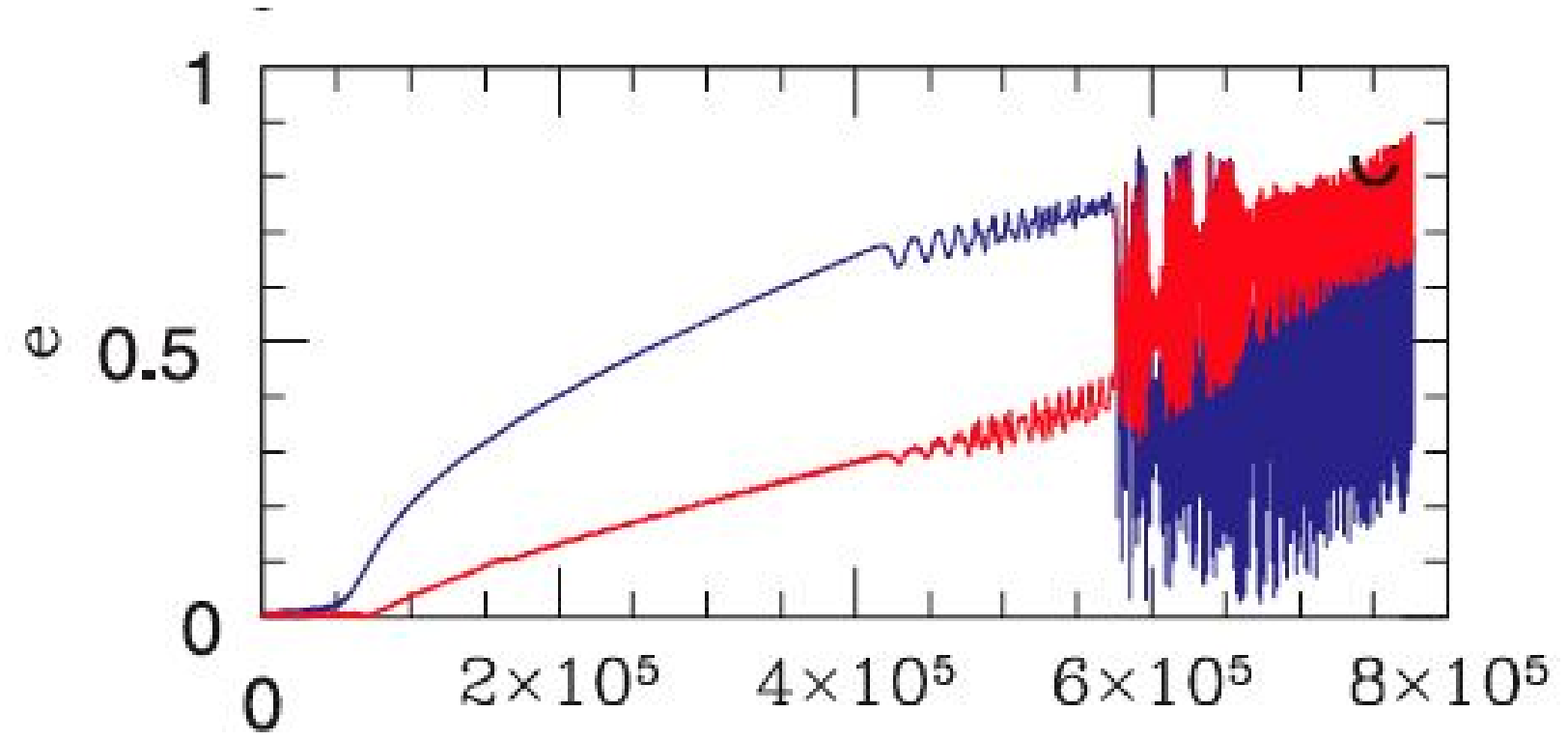}{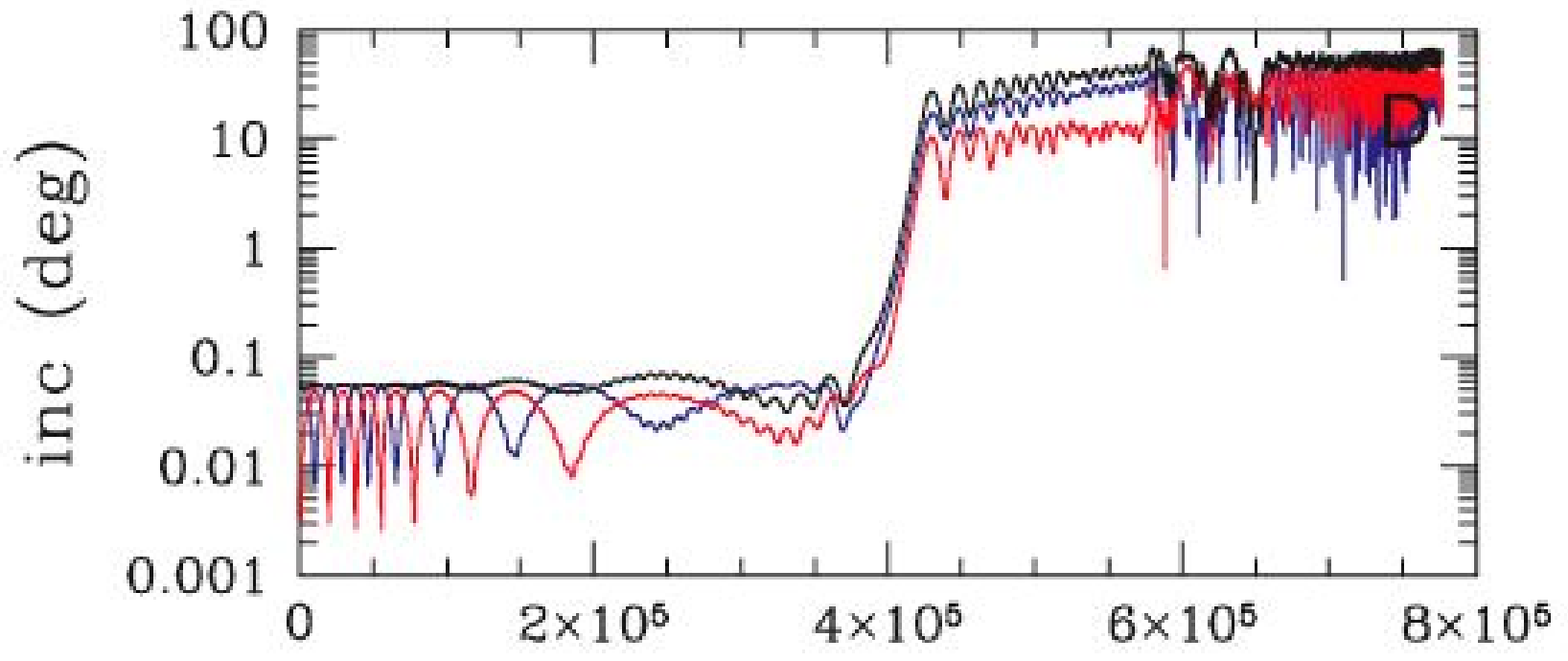}
  \caption{\small 
  Evolution of orbital eccentricity (top) and inclination (bottom) of a two-planet system in which an inward migration
  of $10^{-5} \AU \yi$ is imposed on the outer one. The horizontal
  axis is in units of years. Both planets have a mass equal to Jupiter's mass. The orbital eccentricities
  of the planets grow as a result of resonant migration in the 2:1 resonance.  
  The eccentricity of the inner planet grows faster than the outer one. 
  A small initial mutual
  inclination grows to large values as the eccentricities increase. From {\em Tommes \& Lissauer}, 2003.    }
 \label{f:mig-e-i}
 \end{figure}

To maintain a circular orbit, a migrating planet must experience
energy, $E$, and angular momentum, $J$. changes that satisfy
$\dot{E} = \Omega_{\rm p} \dot{J}$, where  the angular speed of the planet $\Omega_{\rm p}$ varies
as the planet migrates.
 As the planets migrate together in a resonant configuration, their mutual interactions cause
 deviations from this relation.
As a result, their energies and angular momenta evolve
in a way that is  incompatible with maintaining
a circular orbit.  Substantial orbital eccentricities
can develop as a consequence  of migration ({\em Yu \& Tremaine}, 2001; {\em Lee \& Peale}, 2002).
In addition, migration can cause a large amplification of an initially small
mutual inclination of the planetary orbit planes ({\em Yu \& Tremaine}, 2001; {\em Thommes \& Lissauer}, 2003;
see Fig.~\ref{f:mig-e-i}).

Planetary system GJ876 is a well-studied case in which the planets
are in a 2:1 resonance. If the system's  measured eccentricities are due to
resonant migration, then according to theory 
({\em Lee \& Peale}, 2002), the system migrated inward by less than 10\%.
Such a small amount of locked migration seems unlikely.
It is more reasonable to expect that some process limited further
eccentricity growth.
Disk-planet interactions could have limited the eccentricities and inclinations
that could be developed by resonant migration. For a single planet interacting
with a disk, theory and simulations suggest that eccentricity is generally damped
for planets of low mass, too low to open a gap. A higher mass (gap opening) planet can undergo eccentricity
growth due to its interaction with the disk. But the level of eccentricity
produced is limited and eccentricity damping occurs above that level
({\em D'Angelo et al}, 2006). The disk-planet mutual inclination can also be suppressed or
limited by the dissipation of disk warps ({\em Lubow \& Ogilvie}, 2001).
Recent simulations suggest that disk-planet interactions due to gas
interior to the orbit of the inner planet could have limited the eccentricities of multi-planet
systems  to observed levels ({\em Crida et al.} 2008). 
But it may be possible under certain circumstances, such as a depleted inner disk, that large eccentricities
and orbital inclinations could develop due to such processes.

\bigskip
\noindent
\textbf{ 3.3 Other Forms of Migration}
\bigskip

{\it Migration in  a Planetesimal Disk}

After the gaseous disk is cleared from the vicinity of the star,
after about $10^7 y$, there remains a disk of solid material
in the form of low mass planetesimals. This disk is of
much lower mass than the original gaseous disk. But the disk
is believed to have caused some migration in the early
solar system with important
consequences ({\em Hahn \& Malhotra}, 1999; {\em Tsiganis et al}, 2005).

There is strong evidence that Neptune migrated outward. This evidence comes from observations
of Kuiper belt objects that are resonantly trapped
exterior, but not interior, to Neptune's orbit.
The detailed dynamics of a planetesimal disk are somewhat different
from the case of a gaseous disk, as considered in Section 2. 
The planetesimals behave as a nearly collisionless system of particles.
Jupiter  is much more massive
than the other planets and can easily absorb angular momentum changes
in  Neptune. As Neptune scatters planetesimals
inward and outward, it undergoes angular momentum changes.
It is the presence of Jupiter that breaks the symmetry in Neptune's angular
momentum changes. Once an inward scattered planetesimal
reaches the orbit of Jupiter, it gets flung out with considerable
energy and does not interact again with Neptune.
As a result of the loss of inward scattered particles, Neptune
gains angular momentum and migrates outward, while
Jupiter loses angular momentum and migrates slightly inward.

A somewhat analogous process  occurs in gaseous circumbinary disks, disks that orbit around binary 
star systems ({\em Pringle}, 1991). The circumbinary disk gains angular momentum at the expense of the binary. The binary orbit contracts
as the disk outwardly expands. 
A gap-opening planet embedded in a circumbinary
disk (or under some conditions, a disk that surrounds
a star and massive inner planet) would undergo a form of Type~II migration
that could carry the planet outward ({\em Martin et al}, 2007). In the
solar system case, the Sun-Jupiter system plays the role of the binary. 
Viscous torques  are the agent for transferring the angular
momentum from the binary outward in the gaseous circumbinary disk, while particle torques
play the somewhat analogous role in the planetesimal disk.

In a planetesimal disk, another process can operate
to cause migration. This process is  
similar to the runaway
coorbital migration (Type~III migration) described above, but applied
to a collisionless system of particles ({\em Ida et al.} 2000).
Interactions between the planetesimals and the planet in the
planet's coorbital zone can give rise to a migration instability.

{\it Kozai Migration}

A planet that orbits a star  in a binary star system
can periodically undergo a temporary large increase in 
its orbital eccentricity
through the a process known as the Kozai effect (see also Chapter 10). 
Similar Kozai cycles can occur in multi-giant planet systems.
The basic idea behind Kozai migration is that
the increased eccentricity brings the planet closer to the
star where it loses orbital energy through tidal dissipation.
In the process, the planet's semi-major axis is reduced and 
inward migration occurs ({\em Wu \& Murray}, 2003). 
We describe this in more detail below.

Consider a planet in a low eccentricity orbit
that is well interior to the binary orbit and is initially highly inclined 
with respect to it.  The orbital plane of the planet 
can be shown to undergo
tilt oscillations on timescale of $\sim P_{\rm b}^2/P_{\rm p}$,
where $P_{\rm b}$ is the binary orbital period, $P_{\rm p}$ is the
planet's orbital period, and by assumption $P_{\rm b} \gg P_{\rm p}$.   Under such conditions,
it can be shown that the component of the planet's angular
momentum perpendicular to the binary orbit plane (the $z$-component) is approximately
conserved, $J_z = M_{\rm p} \sqrt{G M_{\rm s} a_{\rm p} (1- e_{\rm p}^2) } \cos{I}$, where
$a_{\rm p}$ and
$e_{\rm p}$ are respectively the semi-major axis and
eccentricity of the planet's orbit and $I$ is the inclination
of the orbit with respect to the plane of the binary.

The conservation of $J_z$ is easily seen in the case that the binary orbit is circular and the companion star is of low mass compared to the mass of the
star about which the planet orbits. On such long timescales $ \gg P_{\rm b}$,
the companion star can be considered to be a continuous ring
that provides a static potential. In that case, the azimuthal
symmetry of the binary potential guarantees
that $J_z$ is conserved.
By assumption, we have $\cos{I} \ll 1$ and $e_{\rm p} \ll 1$ in the initial
state of the system.
As the planet's orbital plane evolves and passes into alignment with
the binary orbital plane, $\cos{I} \sim 1$,  conservation of $J_z$
requires $e_{\rm p} \sim 1$. In other words, $J_z$ is initially small 
because of the high inclination of the orbit. When the inclination
drops, the orbit must become more eccentric (radial), 
in order to maintain the same
small $J_z$ value. The process then periodically trades high inclination for
high eccentricity.

During the times of increased eccentricity, the planet
may undergo a close encounter with the central star at periastron
distance $a_{\rm p} (1-e_{\rm p})$. During the encounter, the tidal dissipation
involving the star and planet
results in an energy loss in the orbit of the planet and therefore
a decrease in $a_{\rm p}$. This process then results in inward planet migration.
The energy loss may occur over several oscillations of the orbit plane.

Another requirement for the Kozai process to operate is that
the system must be fairly clean of other bodies. The presence of
another 
object
could induce a precession that washes out the Kozai effect.
Given the special requirements needed for this process to operate,
it is not considered to be the most common form of migration.
However, there is good evidence that it does operate in some
systems ({\em Takeda \& Rasio}, 2005). The Kozai effect can also occur in two-planet systems,
where the outer planet plays the role of the binary companion.
The process can be robust due to the proximity of outer planet
 ({\em Nagasawa et al.} 2008).

\bigskip
\noindent
\textbf{ 3.4  Techniques Used in Numerical Simulations}
\bigskip

Numerical simulations provide an important
tool for analyzing planet migration.  They can provide
important insights in cases where nonlinear and time-dependent
effects are difficult to analyze by analytic methods.
Simulations, as well as analytical models, depend on a model of appropriate physical
processes, such as heating and cooling, the treatment of the
disk turbulence (either by a viscosity or detailed modeling of the disk instability that causes the turbulence),
and the model of the disk structure. Below we discuss some
issues related to the use of simulations for simple (isothermal and alpha-disk)  multi-dimensional
models of disk-planet interactions.

Some powerful grid-based hydrodynamics codes (such as the ZEUS code
{\em Stone \& Norman}, 1992)
 have been 
adapted to the study of disk-planet interactions. 
In addition, particle codes based on the Smoothed Particle Hydrodynamics
(SPH) ({\em Monaghan,} 1992) have sometimes been employed.
A systematic comparison between many of the codes has been
carried out by {\em de Val-Borro et al}, 2006.
We discuss
a few basic points.
For planets that open a gap, grid-based codes
offer an advantage over particle-based codes.
The reason is that the resolution of grid based codes is determined
by the grid spacing, while the resolution of particle-based codes is determined by the particle density. If a planet opens an imperfect gap, 
the particle density and resolution  near the planet  is low.
Higher resolution occurs where the particle density is higher.
On the other hand, SPH well represents regions of a cold disk that are away from a planet.
The gas follows the trajectories of SPH particles that are
slightly modified by gas pressure and tidal forces. SPH
provides high resolution in dense regions,
such as near the cores of low mass planets embedded in disks
({\em Ayliffe \& Bate}, 2007).

There have been variable resolution 
techniques developed for grid-based
codes in which the highest resolution is provided
in regions near the planet where it is needed, such as nested grid methods
({\em D'Angelo et al}, 2002). Such techniques need to
provide a means of joining the regions of high
and low resolution without introducing artifacts (such as wave
reflections) or
lowering the overall accuracy of the scheme.

Grid-based codes that simulate disk-planet interactions 
typically employ numerical devices to improve convergence.
For example, the gravitational potential of the planet is 
often replaced by one that does not diverge near the planet.
The potential is limited by introducing a smoothing length,
a distance 
within which the potential does not increase near the planet.
Another limitation is that the simulated domain of the disk is typically limited to a region
much smaller than the full extent of the disk. Techniques have
been developed to ensure that reflections from the boundaries
do not occur, e.g., by introducing enhanced wave dissipation
near the boundary or approximate outgoing wave boundary conditions.

The time-steps of codes are limited by the Courant condition.
Short time-steps often result from
the region near the inner boundary of the
computational domain (smaller radii) where the disk rotation is fastest.
As a result, it is difficult to extend the disk very close to the
central star. The FARGO scheme ({\em Masset}, 2000) 
is very useful in overcoming this limitation.
However, the method is difficult to apply to a variable grid spaced code.

Convergence is a major issue with these simulations.
Ideally, one should demonstrate that the results of
the simulations are sufficiently insensitive to the locations of the boundaries,
the size of the smoothing length, the size of the time steps, and the grid resolution.
Even the direction of migration can be affected by the size of the potential smoothing
length in certain cases. In a 2D simulation, a finite smoothing length
$\sim H$ provides a means of simulating the reduced
effects of planet gravity on a disk of finite thickness. But in the 2D
case, the limit of zero smoothing length is unphysical.
In practice, testing for convergence is computationally expensive, but can be done
for a subset of the models of interest, perhaps over a limited
time range. Demonstrating
convergence is important for providing reliable results.

\bigskip






\bigskip

\bigskip
\centerline{\textbf{ 4. FUTURE PROSPECTS}}
\bigskip

The theory of planet migration is intertwined with the theory
of planet formation (Chapter 13). The timescale for a
planet to grow within a disk is a key element in understanding
whether planet migration is a major obstacle to planet
formation. We have pointed out that the 
formation timescales for gas giant planets in the core accretion
model do present a problem for the simplest planet migration
theories. However, alternative migration models, some of which are
described in Section 3, may be appropriate. Future prospects
for resolving this issue depend on advances in the theory
of planet formation.

Prospects for making progress in the theory
of planet migration  rely on including and more accurately representing
physical processes in the models, improving computer simulations,
and obtaining a better understanding of the structure of planet-forming gaseous disks.

Planet migration in disks is a consequence of the
action of Lindblad and corotational resonances, as described in Section 2.
The theory of Lindblad resonances is better understood.
Their linear and nonlinear properties have been analyzed in more detail.
The corotational resonances are somewhat more delicate
and less well understood.
Progress on the theory of planet migration will likely involve
further investigations of the role of corotational resonances (e.g., {\em Paardekooper \& Mellema}, 2006).

Improvements to computer capabilities and codes should allow
simulations to be carried out with higher resolution over longer timescales.
Progress will be made by also including 
more physical effects.  For example, most multi-dimensional simulations have made
only the simplest assumptions about the thermal properties
of the disk. The use of a turbulent viscosity is a serious approximation to the effects of turbulence.
Directly simulating the instability that produces the turbulence along with
planet migration is a major computational challenge (e.g., {\em Nelson,} 2005).

Some calculations
have suggested that higher mass eccentric orbit planets could undergo
slowed or outward migration due to their interactions
with a gaseous disk. But we do not know whether such planets
have acquired their eccentricities at this early stage.
Observations of the eccentricities young planets 
would be quite valuable in understanding this issue.
It would be useful to know whether planetary eccentricities are
determined at early times, suggesting that planet eccentricities are present
while the planet is embedded in its gaseous disk.

A major uncertainty in the theory of planet migration is the physical
state of the disk. Low mass planet 
migration behaves very differently depending on the level of disk
turbulence and the structural properties of the disk (Section 3). For example,
in weakly turbulent disks, feedback effects may halt Type~I migration.
While in highly turbulent disks, fluctuating torques may dominate over Type~I torques.  
The presence of rapid radial density
variations or weakened (or inverted) radial temperature variations
can substantially alter migration, since it depends
on the competition between torques involving material
just inside and outside the orbit of the planet.
The better determination of disk properties
will likely rely on some combination of improved theory and observations. 
It is unlikely that theory alone will be able to make
much progress along these lines because, for example, the density
structure depends on the disk turbulence that is difficult to accurately predict
from first principles. Once such difficulty is that 
magnetically driven turbulence is strongly affected by
the abundance and size distribution of dust grains that control the level of disk
ionization (e.g., {\em Salmeron \& Wardle, 2008}). 
The observational determination of disk properties (disk density and temperature as a function of distance
from the star)
in the inner parts of protostellar
disks, where planet formation is expected to occur, is important for such purposes.
New telescopes such as ALMA and JWST may be quite
valuable in making such determinations.


\bigskip
\textbf{ Acknowledgments.} This work was partially supported by 
NASA grant NNX07AI72G to SL. We thank Phil Armitage, Gennaro D'Angelo and Jim Pringle
for carefully reading a draft and suggesting improvements. We thank an anonymous
reviewer for many helpful suggestions. We also benefited considerably by discussions
at the Cambridge University Isaac Newton Institute Program  "Dynamics of Discs and Planets."

\bigskip

\centerline\textbf{ REFERENCES}
\bigskip
\parskip=0pt
{\small
\baselineskip=11pt

\refs Alibert, Y., Mousis, O., Mordasini, C. and Benz, W. (2005)
New Jupiter and Saturn Formation Models Meet Observations.
{\em Astrophys. J., 626}, L57-L60. 

\refs 
Armitage, P.~J., 
Livio, M., \& Pringle, J.~E.\ (2001), 
{\em Mon.\ Not.\ Roy.\ Astr.\ Soc., 324}, 705-711.

\refs Artymowicz, P. (1993)	
Disk-Satellite Interaction via Density Waves and the Eccentricity Evolution of Bodies Embedded in Disks. 
{\em Astrophys. J., 419}, 166-180.

\refs Artymowicz, P. and Lubow, S. H. (1996)	
Mass flow through gaps in circumbinary disks. {\em Astrophys. J., 467}, L77.

\refs Artymowicz, P. (2004)
Migration Type III.
{\it KITP Conference on Planet Formation,}
http://online.kitp.ucsb.edu/online/planetf\_c04

\refs Ayliffe, B. A. and Bate, M. R. (2009)	
Gas accretion on to planetary cores: three-dimensional self-gravitating radiation hydrodynamical calculations. 
{\em Mon.\ Not.\ Roy.\ Astr.\ Soc., 393}, 49-64.

\refs Balbus, S. A., and Hawley, J. F. (1991) A powerful local shear instability 
in weakly magnetized disks. I. Linear analysis. {\em Astrophys. J., 376}, 
214-222. 

\refs  Baruteau, C. and Masset, F. (2008)
On the Corotation Torque in a Radiatively Inefficient Disk. {\em Astrophys. J., 672}, 
1054-1067.

\refs Bate, M.~R., Lubow, S.~H.,  Ogilvie, G.~I., and Miller, K.~A. (2003)
Three-dimensional calculations of high- and low-mass planets embedded in protoplanetary discs.
 {\em Mon.\ Not.\ Roy.\ Astr.\ Soc., 341}, 213-229.

\refs Bodenheimer, P., 
Hubickyj, O., and Lissauer, J.~J.  (2000)  
Models of the in Situ Formation of Detected Extrasolar Giant Planets. {\em Icarus}, 143, 2. 

\refs Butler, R.~P., et al. (2006)
Catalog of Nearby Exoplanet.
 {\em Astrophys. J., 646}, 505-522.

\refs Crida, A., Morbidelli,
A., and Masset, F.  (2006),
On the width and shape of gaps in protoplanetary disks.
{\em Icarus, 181}, 587.

\refs Crida, A., S{\'a}ndor, Z., \& Kley, W.\ (2008),
Influence of an inner disc on the orbital evolution of massive planets migrating in resonance,
 {\em Astron, Astrophys., 483}, 325-337. 

\refs Cuzzi, J.~N., and Weidenschilling, S.~J. (2006), 
Particle-Gas Dynamics and Primary Accretion,
in {\em Meteorites and the Early Solar System II},  (Lauretta \& McSween Jr. eds.), Univ. Arizona Press, Tucson.

 \refs D'Angelo, G., Henning, T.,  and  Kley, W. (2002)
 Nested-grid calculations of disk-planet interaction.
 {\em Astron, Astrophys., 385}, 647-670. 

\refs D'Angelo, G., Lubow, S.~H., and Bate, M.~R. (2006)
Evolution of Giant Planets in Eccentric Disks.
 {\em Astrophys. J., 652}, 1698-1714.

\refs D'Angelo, G., and Lubow, S.~H. (2008)
Evolution of Migrating Planets Undergoing Gas Accretion. 
 {\em Astrophys. J., 685}, 560-583. 
 
\refs Gammie, C. F. (1996) Layered accretion in T Tauri disks. {\em Astro- 
phys. J.}, 457, 355Ð362.


\refs Goldreich, P., and Tremaine, S. (1979)
The Excitation of Density Waves At the Lindblad and Corotation
Resonances By an External Potential
{\em Astrophys. J., 233}, 857-871.

\refs Goldreich, P., and Tremaine, S. (1980)
Disk-satellite interactions.
{\em Astrophys. J., 241}, 425-441.

\refs Goldreich, P., and Tremaine, S. (1982)
The Dynamics of Planetary Rings.
{\em Ann. Rev. Astron. Astrophys., 20}, 249-283.

\refs Goldreich, P., and Sari, R. (2003)
Eccentricity Evolution For Planets in Gaseous Disks.
{\em Astrophys. J., 585},  1024Ð1037.

\refs Hahn, J.~M., \& Malhotra, R. (1999)
Orbital Evolution of Planets Embedded in a Planetesimal Disk. 
 {\em Astron. J., 117}, 3041-3053.

\refs Hayashi, C. (1981) Structure of the solar nebula, growth and de- 
cay of magnetic Þelds and effects of magnetic and turbulent 
viscosities on the nebula. {\em Progress of Theoretical Physics Sup- 
plement, 70}, 35Ð53. 

\refs Hourigan, K. \& Ward, W.~R. (1984) Radial Migration of Preplanetary
Material: Implications for the
Accretion Time Scale Problem, {\em Icarus, 60}, 29-39.

\refs Ida, S., Bryden, G., Lin, D. N. C. and Tanaka, H. (2000)
Orbital migration of Neptune and orbital distribution of trans-Neptunian 
objects. {\em Astrophys. J., 534}, 428-445. 

\refs Ida, S., and Lin, D. N. C. (2004) Toward a deterministic model of 
planetary formation. I. A desert in the mass and semimajor axis 
distributions of extrasolar planets.
{\em Astrophys. J., 604}, 388-413. 

\refs Ida, S., Guillot, T., and Morbidelli, A. (2008) 
Accretion and destruction of planetesimals in turbulent disks.
{\em Astrophys. J., 686}, 1292-1301.

\refs Ida, S., and Lin, D. N. C. (2008) Toward a deterministic model of 
planetary formation. IV. Effects of type I migration.
{\em Astrophys. J., 673}, 487-501. 

\refs Jang-Condell, H., and Sasselov, D. D.  (2005) 
Type~I Planet Migration in a Nonisothermal Disk.
{\em Astrophys. J., 619}, 1123Ð1131.

\refs Johnson, E.~T.,  Goodman, J., and Menou, K. (2006)
Diffusive Migration of Low-Mass Protoplanets in Turbulent Disks.
{\em Astrophys. J., 647}, 1413-1425.


\refs Lee, M.~H., and Peale, S.~J. (2002)
Dynamics and Origin of the 2:1 Orbital Resonances of the GJ 876 Planets.
 {\em  Astrophys. J., 567}, 596-609.

\refs Li,  H., Lubow, S.~H., Li, S., and Lin,  D.~N.~C. (2009)
Type I Planet Migration in Nearly Laminar Disks.
 {\em  Astrophys. J., 690}, L52.

\refs Lin, D.~N.~C., and Papaloizou, J. (1979)
Tidal torques on accretion discs in binary systems with extreme mass ratios.
 {\em Mon.\ Not.\ Roy.\ Astr.\ Soc., 186}, 799-812.

\refs Lin, D.~N.~C., and Papaloizou, J.~C.~B. (1986)
On the tidal interaction between protoplanets and the protoplanetary disk. III - Orbital migration of protoplanets.
 {\em  Astrophys. J., 309}, 846-857.
 
 \refs Lubow, S. H. (1991)
A model for tidally driven eccentric instabilities in fluid disks.
 {\em  Astrophys. J., 381}, 259-267.
 
 \refs Lubow, S. H. and Ogilvie, G. I. (2001)	
Secular Interactions Between Inclined Planets and a Gaseous Disk. {\em Astrophys. J., 560}, 997-1009.

\refs Lubow, S. H. and D'Angelo, G. (2006)	
Gas Flow Across Gaps in Protoplanetary Disks. {\em Astrophys. J., 641}, 526.

\refs Lynden-Bell, D. and Pringle, J. E. (1974)
The Evolution of Viscous Discs and the Origin of the Nebular Variables.
 {\em  Mon.\ Not.\ Roy.\ Astr.\ Soc., 168}, 603-637.

 \refs Martin, R.~G., Lubow, S.~H., Pringle, J.~E., \& Wyatt, M.~C. (2007)
Planetary migration to large radii.
{\em  Mon.\ Not.\ Roy.\ Astr.\ Soc., 378}, 1589-1600.

\refs Masset, F.~S. (2000)
FARGO: A fast eulerian transport algorithm for differentially rotating disks.
{\em Astron. Astrophys. Supp., 141}, 165-173.

\refs Masset, F.~S.,  and Papaloizou, J.~C.B. (2003)
Runaway migration and the formation of hot Jupiters .
{\em Astrophys. J., 588}, 494Ð508.

\refs Masset, F.~S., Morbidelli, A., Crida,  A., and Ferreira, J. (2006)
Disk surface density transitions as protoplanet traps.
{\em Astrophys. J., 642}, 478-487.

\refs Matsumura, S., Pudritz, R.~E. and Thommes, E.~W. (2007)
Saving Planetary Systems: Dead Zones and Planetary Migration,
{\em Astrophys. J., 660}, 1609-1623.

\refs Mayor, M., and Queloz, D. (1995) A Jupiter-mass companion to a solar-type 
star. {\em Nature, 378}, 355-359.

\refs  Menou, K., and Goodman, J. (2004)
Low-Mass Protoplanet Migration in T Tauri Disks.
{\em Astrophys. J., 606}, 520-531.

\refs Monaghan, J.~J. (1992)
Smoothed particle hydrodynamics.
{\em Ann. Rev. Astron. Astroph., 30}, 543-574.

\refs Nagasawa, M., Ida, S., and Bessho, T. (2008)	
Formation of hot planets by a combination of planet scattering, 
tidal circularization, and the Kozai mechanism.
{\em Astrophys. J., 678}, 498-508.

\refs Nelson, R.~P.  (2005)
On the orbital evolution of low mass protoplanets in turbulent, magnetised disks.
 {\em  Astron. Astrophys. 443}, 1067-1085.

\refs Ogilvie, G.~I., and Lubow, S.~H. (2003)
Saturation of the Corotation Resonance in a Gaseous Disk.
{\em Astrophys. J., 587}, 398-406.

\refs Ogilvie, G.~I., and Lubow, S.~H. (2006)
The effect of planetary migration on the corotation resonance.
{\em  Mon.\ Not.\ Roy.\ Astr.\ Soc., 370}, 784-798.

\refs Paczynski, B. (1978)
A model of selfgravitating accretion disk.
 {\em Acta Astronomica, 28}, 91 

\refs  Paardekooper, S.-J., and Mellema, G. (2006)
Halting type I planet migration in non-isothermal disks.
 {\em  Astron. Astrophys. 459}, L17-L20.
 
 \refs Papaloizou, J.~C.~B. (2002)
 Global m = 1 modes and migration of protoplanetary cores in eccentric protoplanetary discs.
  {\em  Astron. Astrophys. 388},  615-631.
 
 \refs  Pringle, J.~E. (1991)
 The properties of external accretion discs.
 {\em Mon.\ Not.\ Roy.\ Astr.\ Soc., 248}, 754-759. 

\refs  Rafikov, R.~R. (2002)
Planet Migration and Gap Formation by Tidally Induced Shocks.
{\em Astrophys. J., 572}, 566-579.

\refs  Rice, W. K. M and Armitage, P. J. (2005)
Quantifying Orbital Migration From Exoplanet Statistics and Host Metallicities.
{\em Astrophys. J., 630}, 1107-1113.

\refs Stone, J.~M., and Norman, M.~L. (2002)
ZEUS-2D: A radiation magnetohydrodynamics code for astrophysical flows in two space dimensions. I - The hydrodynamic algorithms and tests.
{\em Astrophys. J. Supp., 80}, 753-790.

\refs Salmeron, R., and Wardle, M.\ (2008) 
Magnetorotational instability in protoplanetary discs: the effect of dust grains.
{\em Mon.\ Not.\ Roy.\ Astr.\ Soc.,  388}, 1223-1238.

\refs Shakura N.~I., and Sunyaev R. A., (1973), Black holes in bi- 
nary systems. Observational appearance.
{\em Astron. Astrophys. 24}, 337 - 355.


\refs Takeda, G., and Rasio, F. A. (2005)
High Orbital Eccentricities of Extrasolar Planets Induced by the Kozai Mechanism.
{\em Astrophys. J., 627}, 1001-1010.

\refs  Tanaka, H., Takeuchi, T., \& Ward, W.~R. (2002)
Three-Dimensional Interaction between a Planet and an Isothermal Gaseous Disk. I. Corotation and Lindblad Torques and Planet Migration.
{\em Astrophys. J., 565}, 1257-1274.

\refs Terquem,  C.~E.~J.~M.~L.~J. (2003)
Stopping inward planetary migration by a tooroidal magnetic field.
{\em  Mon.\ Not.\ Roy.\ Astr.\ Soc., 341}, 1157-1173.

\refs Terquem, C.~E.~J.~M.~L.~J.\ 
(2008), New composite models of partially ionized protoplanetary disks, {\em Astrophys. J., 689}, 532-538

\refs Thommes, E.~W., and Lissauer, J.~J. (2003)
Resonant Inclination Excitation of Migrating Giant Planets.
{\em Astrophys. J., 597}, 566-580.

\refs Tsiganis, K., Gomes,  R., Morbidelli, A., \& Levison, H.~F. (2005)
Origin of the orbital architecture of the giant planets of the Solar System.
{\em Nature, 435}, 459-461.

\refs de Val-Borro, M., et al. (2006)
A comparative study of disc-planet interaction.
{\em Mon.\ Not.\ Roy.\ Astr.\ Soc., 370}, 529-558.

\refs Ward, W.~R. (1991)
Horeshoe Orbit Drag.
{\em Lunar and 
Planetary Institute Conference Abstracts, 22}, 1463-1464.

\refs Ward, W.~R. (1997)
Protoplanet Migration by Nebula Tides.
{\em Icarus, 126}, 261-281.

\refs Weidenschilling, S. J. (1977) Aerodynamics of solid bodies in the solar nebula. 
{\em Mon.\ Not.\ Roy.\ Astr.\ Soc., 180}, 57-70. 

\refs Wu, Y., and Murray, N. (2003)
Planet Migration and Binary Companions: The Case of HD 80606b.
{\em Astrophys. J., 589}, 605-614.

\refs Yang, C.~C., Mac Low, M.-M., and Menou, K. (2009)
Planetesimal and Protoplanet Dynamics in a Turbulent Protoplanetary Disk:
Ideal Unstratified Disks,
{\em astro-ph}, arXiv:0907.1897v1.

\refs Yu, Q., and Tremaine, S. (2001)
Resonant Capture by Inward-migrating Planets
{\em Astron. J., 121}, 1736-1740.

\end{document}